# Leveraging Discarded Samples
# for Tighter Estimation of Multiple-Set Aggregates

Edith Cohen
AT&T Labs–Research
Florham Park, NJ 07932, USA
edith@research.att.com

Haim Kaplan
Tel Aviv University
Tel Aviv, Israel
haimk@cs.tau.ac.il

## Abstract

*Many datasets such as market basket data, text or hypertext documents, and sensor observations recorded in different locations or time periods, are modeled as a collection of sets over a ground set of keys. We are interested in basic aggregates such as the weight or selectivity of keys that satisfy some selection predicate defined over keys' attributes and membership in particular sets. This general formulation includes basic aggregates such as the Jaccard coefficient, Hamming distance, and association rules.*

*On massive data sets, exact computation can be inefficient or infeasible. Sketches based on coordinated random samples are classic summaries that support approximate query processing. Queries are resolved by generating a sketch (sample) of the union of sets used in the predicate from the sketches these sets and then applying an estimator to this union-sketch.*

*We derive novel tighter (unbiased) estimators that leverage sampled keys that are present in the union of applicable sketches but excluded from the union sketch. We establish analytically that our estimators dominate estimators applied to the union-sketch for* all *queries and data sets. Empirical evaluation on synthetic and real data reveals that on typical applications we can expect a 25%-4 fold reduction in estimation error.*

## 1. Introduction

We consider datasets modeled as a collection $\mathcal{S}$ of (possibly intersecting) *sets*, defined over a ground set $I$ of (possibly weighted) *keys*. A classic example is documents over features or terms, according to presence in the document.

Basic aggregates over such data are *weight* and *selectivity* of subpopulations of keys. A query specifies a *subpopulation* of $I$ by a selection predicate. The weight aggregate is the sum of the weights of the keys that satisfy the predicate. If keys have uniform weights, the weight aggregate is known as DV (distinct values) count. An example of a weight query is the number of terms present both in document $A$ and in document $B$ and are at least 5 characters long. Selectivity queries are defined with respect to some (sub) collection of sets: The result is the ratio of the sum of the weights of all keys in the union of these sets for which the predicate holds and the total weight of the union of these sets. An important selectivity aggregate is the *Jaccard coefficient* of $A$ and $B$ defined as $|A \cap B|/|A \cup B|$, which measures the similarity between $A$ and $B$. A common technique to enhance this similarity metric is to assign larger weights to features/terms that are less frequent in the corpus. For weighted keys, the Jaccard coefficient generalizes to $w(A \cap B)/w(A \cup B)$ (the ratio of the weight of the intersection and the weight of the union).

Basic (approximate) weight aggregates are also used to compute more complex (approximate) aggregates, such as variance [15] of a subpopulation of keys or ratio of the weights of two subpopulations of keys.

The selection predicates that specify subpopulations are defined using conditions on keys' attributes *and* on keys' memberships in the different sets. We distinguish between *attribute-based* conditions, that are based on properties available through the identifier of the key (length, origin, or frequency of a term, type of feature) and *membership-based* conditions that are based on the key's set memberships. For example, terms common to two documents $A, B$ are specified using the predicate with membership-based conditions "in A and in B". The predicate "in A and not in B and length $\geq$ 5" has both attribute-based (length of a term) and membership-based conditions.

We list additional datasets that fall in this framework.

- **Sensor nodes recording daily vehicle traffic in different locations in a city:** Keys are distinct vehicles (license plate numbers) and sets are location-date pairs (all vehicles observed at that location that date). Example queries with membership-based conditions: "number of distinct vehicles which operated in Manhattan on election day, 2008" (size of the union of all Manhattan locations in election day) or "number of distinct vehicles operated in Tribeca on both Sunday and Monday on election week" (size of intersection of the union of locations in Tribeca neighborhood in Monday and Tuesday); "number of vehicles that crossed both the Hudson and the East River on Independence day 2008" (size of intersection of the union of bridges/tunnels across the Hudson and bridges/tunnels across the East River) etc. Queries can be re-stricted to particular classes of vehicles (e.g., taxi cubs or heavy trucks) by adding attribute-based conditions. Such queries can be used for planning purposes.

• **Market-basket dataset:** Keys are goods, each with an associated marketing cost (these are the weights). Each customer (basket) defines a set which is the set of goods she purchased. Example queries are "the total marketing cost of baby products purchased by male customers from Union county." This predicate has attribute-based condition (product type) and membership-based conditions (specification of the customer segment as a union of sets).

• **"Inverted" market-basket dataset:** Keys are baskets (customers) and sets are goods (all baskets containing that particular good). A query that asks "what is the likelihood that a certain item is purchased given that another item is purchased" (this is an "association rule" [1, 42]) can be expressed using a predicate with membership-based conditions. If $A$ is the set of customers purchasing, say beer, and $B$ is the set of customers purchasing diapers then the selectivity of $A\cap B$ with respect to $B$ is just the likelihood that a person purchases beer given that she/he purchased diapers. This query can be narrowed down to a particular customer segment (eg, by zip code or gender) if we add an attribute-based conditions to the predicate.

• **Hyperlinked documents**: Sets and keys are documents, where the set of document $A$ includes all documents with hyperlinks to document $A$. Documents may be weighted by access data or page rank. Example queries are "the total weight of documents referencing at least 5 out of the 10 documents in $Q$." This predicate has membership-based conditions.

• **P2P network:** Keys are files and sets are all neighborhoods of all peers (sets of files shared by peers in that neighborhood). Example queries are "the weight of files stored in the 5-hop neighborhoods of peer $A$ or peer $B$," or "number of distinct files in a particular subject in the 3-hop neighborhood of peer $A$." Such queries can be used to keep the search focused on peers that contains many keys in a particular topic or peers that are more similar to the querying peer [14, 52, 55].

Exact computation of such queries requires retrieving the full content of all sets relevant to the predicate, computing the union, and applying the predicate to all keys in the union, adding up the weights of keys that satisfy the predicate. On massive or distributed data, the high cost of exact computation prohibits running a large number of queries (that is required for clustering or association rule mining). In some cases, such as network traffic data, the full data set may no longer be available at the time the query is formulated.

The practical solution is to produce a summary that supports approximate processing of such queries. A suitable summary format is a set of sketches, one for each set.

A basic sketch format for a single set is a weighted random sample without replacement of the keys in the set, obtained using *order* (*bottom-k*) sampling [46, 12, 47, 7, 42, 51, 17, 26, 19, 2, 32]. The sample is obtained by assigning a random rank value to each key and including the $k$ keys with smallest rank values. The rank distribution for a key depends on its weight. Using different distributions, order samples can realize classic weighted sampling [46, 33] where keys are successively drawn proportionally to their weight or *priority sampling* (*sequential poisson sampling*) [44, 45, 47, 26], which have estimators [26] that (nearly) minimize the sum of per-key variances [53]. The sample of a set (with some auxiliary rank information) supports tight unbiased estimators for weight and selectivity aggregates over the set [19, 17, 26].

Multiple-set aggregates are estimated using the *union-sketch reduction* to estimators over a single-set. The reduction applies when sketches of different sets are *coordinated*, that is, the same set of rank values of keys is used across sets. It is known that without coordination (independently sampling each set), it is not possible to obtain strong estimators [9].

For a multiple-set aggregate and selection predicate with relevant sets $\mathcal{S} \in \mathcal{A}$, a size-$k$ sketch of the union $\bigcup_{A\in\mathcal{S}} A$ is constructed from size-$k$ sketches of the sets in $\mathcal{S}$ [12, 7, 6]. A "single set" weight or selectivity estimator can then be applied to estimate our multiple-set aggregate, by applying it to the union sketch of $\mathcal{S}$.

Coordinated bottom-$k$ sketches can be computed efficiently for diverse data sources including centralized or distributed with *explicitly* or *implicitly* represented sets [17]. Sets are explicitly represented when the data source can be modeled as a list of keys for each set or a list of sets for each key (the inverted data). In the former case, random hash functions are used to decouple the sampling of different sets [7, 8, 24, 6, 42, 2]. Examples of explicitly-represented sets includes item-basket associations in a market basket data, links in web pages, and features in documents [7, 3, 42, 51, 2].

Sets are implicitly represented when memberships are specified indirectly (as in our p2p example) through some metric on a set of points. Implicit representation can be more concise than the corresponding explicit representation. Keys are associated with points and sets are specified by a point and distance pair (*neighborhood*) and include all keys within that distance from the point [12, 22, 21, 41, 36, 16]. Examples are nodes in a graph with the shortest path or reachability metric, the Euclidean plane, or time stamps or sequence numbers on a data stream [12, 21, 16, 36]. When we are interested in multiple distances (neighborhoods) of a point (applications include aggregates with time or spatial decay [21, 16]), *all-distances sketches* succinctly represent coordinated sketches of *all* neighborhoods of the point and can be computed efficiently over the implicit representation of the dataset [12, 16, 17].

**Our contributions.** Our main contribution is the derivation of tighter unbiased estimators for multiple-sets weight and selectivity aggregates. These estimators are applicable to a set of coordinated sketches and therefore apply to the *same* set of sketches as the union-sketch method. We will show that they involve *similar computational tasks* as the union-sketch method and they dominate all previous methods in terms of estimation quality.

**Combinations of sketches.** A close look at the union-sketch approach reveals that we discard potentially useful information present in the *union of the size-$k$ sketches of the sets* by restricting our attention to the *size-$k$ sketch of the union of these sets*. If there are $t$ relevant sets, the union of the sketches includes at least $k$ but up to $t * k$ distinct keys. In Section 3 we consider two more inclusive *combinations* of the sketches of the sets than the union-sketch: The *short combination of sketches* (SCS), and the *long combination of sketches* (LCS). The LCS includes all keys in the union of the sketches, it contains the SCS, which contains the $k$ keys in the sketch of the union.

**Combination** RC **estimators for weight aggregates.** In Section 4 we develop unbiased estimators for subpopulation weight that leverage the additional keys contained in the LCS and SCS. Fully exploiting this additional information was a subtle and chal-lenging task: The SCS can be viewed as a variable-size sequential sample of the union where the number of included keys depends on set memberships of previously selected keys. The LCS can not be expressed as a sequential weighted sample of a set. The challenge lies in benefiting from additional keys without introducing bias – we can not simply apply a single-sketch estimator to combinations.

We build on the powerful Rank Conditioning (RC) estimators that are the best known estimators applicable to a sketch of a single set [19, 26].[1] *Adjusted weights* are assigned to sampled keys and the weight estimate of a subpopulation is the sum of the adjusted weights of sampled keys that belong to the subpopulation. For multiple-set aggregates, our combination RC estimators assign positive adjusted weights to all keys in the combination whereas the basic union-sketch method assigns them only to $k$ keys. We prove that our estimators are unbiased for every subpopulation and furthermore, the covariance of the adjusted weights of any two different keys is zero. This guarantees that the variance of our estimate for a subpopulation is not larger than the sum of the variances of the adjusted weights of the keys in the subpopulation.

We prove that (for *any* selection predicate and data set) the SCS RC estimators are at least as tight (at most the variance) as the union sketch RC estimators. Similarly to union-sketch estimators, the SCS estimators are applicable to general *select* predicates. The LCS RC estimators are at least as tight as the SCS RC estimators but are applicable to a more limited class of predicates that are attribute-based selections from a union of sets. Therefore, our SCS RC estimator strictly dominates all union-sketch based estimators, and for applicable *select* predicates, LCS RC dominates all other methods. In Section 5 we demonstrate how the different estimators are applied.

**Coordinated Poisson samples.** In Section 6 we consider coordinated sketches based on Poisson samples. Poisson sampling (inclusion probabilities of different keys are independent) has the disadvantages over bottom-$k$ sampling of variable sample size and that coordinated Poisson samples can not be computed in a scalable way over implicitly represented sets. We consider estimators for multiple-aggregates that generalize ones proposed in [30, 31] (for uniform weights) and discuss their relation to our bottom-$k$ estimators.

**Getting more from combinations.** Other estimators traditionally applied to the union-sketch can be extended to yield tighter results on combinations:

⋄ **Unbiased selectivity.** In Section 7 we derive unbiased estimators for selectivity queries with respect to the union of the sets in $\mathcal{S}$. While selectivity can be estimated using the ratio of the estimated weight of the set and the estimated weight of the union $\bigcup_{A \in \mathcal{S}} A$, this estimator might be *biased* even if we use unbiased weight estimators. We derive SCS unbiased selectivity estimators that strictly improve over traditional unbiased estimators for Jaccard similarity [7, 6].

⋄ **Maximum Likelihood** (ML). In Section 8 we derive ML estimators applicable to combinations of bottom-$k$ sketches based on successive weighted sampling [46]. The derivation builds on ML estimators [19], and as with other ML estimators, our new ones are biased. We design tailored tighter estimators for applications where the weight of each sketched set is readily available.

**Empirical evaluation.** Section 9 summarizes results of extensive experiments on real and synthetic data. We quantify the power of SCS and LCS-based estimators compared to estimators applied to the union-sketch. Synthetic data was designed to study how performance depends on the relations between the sets and on the number of sets used in the predicate. Real data allowed us to use natural selection predicates and demonstrate potential applications. We discuss related work in Section 10 and conclude in Section 11.

## 2. Preliminaries

This section provides necessary background and definitions.

A *weighted set* $(I, w)$ consists of a set of keys $I$ and a weight function $w$ assigning a $w(i) \geq 0$ to each key $i \in I$. A *rank assignment* maps each key $i$ to a random rank $r(i)$. The ranks of keys are drawn independently using a family of distributions $\mathbf{f_w}$, where the rank of a key with weight $w(i)$ is drawn according to $\mathbf{f_{w(i)}}$. For a set $J$ and a rank assignment $r$ we denote by $r_i(J)$ the $i$th smallest rank of a key in $J$, we also abbreviate and write $r(J) = r_1(J)$. Random rank assignments are used to obtain *sketches* (samples with some auxiliary information) of sets as follows.

The $k$-*mins sketch* [12, 7] of a set $J$ is produced from $k$ independent rank assignments, $r^{(1)}, \ldots, r^{(k)}$. The sketch of a set $J$ is the $k$-vector $(r^{(1)}(J), r^{(2)}(J), \ldots, r^{(k)}(J))$. Depending on the application we may store with each of these ranks, attributes associated with the corresponding key.

A *bottom-$k$ sketch* (or *order sample*) [46, 12, 47, 7, 42, 51, 17, 19, 2, 32] of a set $J$ is defined based on a single rank assignment $r$ as follows. Let $i_1, \ldots, i_k$ be the $k$ keys of smallest ranks in $J$. The sketch consists of $k$ pairs $(r(i_j), w(i_j))$, $j = 1, \ldots, k$, and $r_{k+1}(J)$. (If $|J| \leq k$ we store only $|J|$ pairs.) We denote a bottom-$k$ sketch of a set $A$ with respect to a rank assignment $r$ by $s_k(A, r)$.

Consider a set $\mathcal{A}$ of sets over a set of keys $I$. *Coordinated* $k$-mins or bottom-$k$ sketches are obtained by using the same rank assignment over $I$ (for $k$-mins sketches, same set of rank assignments), when producing the sketches of all sets $A \in \mathcal{A}$. Coordinated sketches should include all rank values and keys' weights. (If we are only interested in predicates with membership-based conditions, then we do not have to include key attribute values in the sketches.)

**The union-sketch.** Coordinated bottom-$k$ and $k$-mins sketches have the property that for a set $\mathcal{S} \subset \mathcal{A}$ of sets we can compute the sketch of $\bigcup_{A \in \mathcal{S}} A$ from the sketches of the sets $A \in \mathcal{S}$. For $k$-mins sketches, the sketch of the union contains, for each rank function the key with minimum rank value across sets in $\mathcal{S}$. For bottom-$k$ sketches, the keys in $s_k(\bigcup_{A \in \mathcal{S}} A, r)$ are the keys with $k$ smallest ranks in $\bigcup_{A \in \mathcal{S}} s_k(A, r)$. Note that $r_{k+1}(\bigcup_{A \in \mathcal{S}} A)$ is the minimum rank of a key that is among the $(k+1)$ smallest ranks in at least one of $A \in \mathcal{S}$ but is not among the $k$ smallest ranks in the union sketch. Therefore, $r_{k+1}(\bigcup_{A \in \mathcal{S}} A)$ can also be determined from the sketches of $A \in \mathcal{S}$.

The union-sketch reduction is a method that allows us to apply a weight/selectivity estimator designed for attribute-based *select* predicates over a single ($k$-mins or bottom-$k$) sketch to estimate the weight/selectivity of a subpopulation specified by a general *select* predicate (with membership and attribute based conditions) over coordinated ($k$-mins or bottom-$k$) sketches of a collection of sets $\mathcal{S}$.

---

[1] There are tighter estimators when the exact total weight of the set is known, but this is not the case in our multiple-set aggregates since the weight of the union of sets can not be exactly recovered from sketches of the sets.

We first identify all sets $\mathcal{S}$ relevant to the predicate. We retrieve the sketches of $\mathcal{S}$ and compute the sketch of the union. A very handy property of the union-sketch is that for each key $x$ included in the sketch of the union we can determine which sets of $\mathcal{S}$ it is a member of. We therefore can treat each membership in a set in $\mathcal{S}$ as an attribute of the keys. We then apply our single-sketch estimator to the union-sketch of $\mathcal{S}$, treating membership-based conditions of the predicate as attribute-based conditions over the keys in the union-sketch.

As a concrete example, consider the inverted market-basket data set and the query "the number of baskets of at most 10 keys that contain beer or wine and cheese." To do so, we isolate the sketches of beer, wine, and cheese, and compute the union sketch. The union sketch is a random sample from the set of baskets that have beer, wine, or cheese. For each basket in the union we know if it has or does not have each one of the three goods. The size of the basket is an attribute. We can therefore identify all baskets in the sample for which the predicate "has beer or has wine and has cheese and has size $\leq 10$" holds and estimate the distinct count.

WS **sketches.** The choice of which family of random rank functions to use matters only when keys are weighted. Otherwise, sketches produced using one rank function can be transformed to any other rank function. Rank functions $\mathbf{f_w}$ with some convenient properties are exponential distributions with parameter $w$ [46, 33, 12]. The density function of this distribution is $\mathbf{f_w(x) = we^{-wx}}$, and its cumulative distribution function is $\mathbf{F_w(x) = 1 - e^{-wx}}$. Equivalently, if $u \in U[0,1]$ then $-\ln(u)/w$ is an exponential random variable with parameter $w$. A useful property for designing estimators [12, 16, 17, 19] is that the minimum rank $r(J) = \min_{i \in J} r(i)$ of a key in a set $J \subset I$ is exponentially distributed with parameter $w(J) = \sum_{i \in J} w(i)$ (the minimum of independent exponentially distributed random variables is exponentially distributed with parameter equal to the sum of the parameters of these distributions).

Moreover, the probability that a key $x \in J$ is the minimum rank key is $w(x)/w(J)$. Hence, a $k$-mins sketch of a set $J$ is a weighted random sample of size $k$, drawn **with replacement** from $J$. We call a $k$-mins sketch using exponential ranks a WSR sketch. On the other hand, a bottom-$k$ sketch of a set $J$ with exponential ranks corresponds to a weighted $k$-sample drawn **without replacement** from $J$ [46, 33]. We call such a sketch a WS sketch.

PRI **sketches.** If the rank value of a key with weight $w$ is selected uniformly at random from $[0, 1/w]$ then the bottom-$k$ sketch is a *priority sketch* (also known as *Sequential Poisson Sample*) [44, 45, 47, 26]. This is the equivalent to choosing rank value $u/w$, where $u \in U[0,1]$ or using density function $\mathbf{f_w(x) = w}$ for $0 \leq x \leq 1/w$ and $\mathbf{f_w(x) = 0}$ otherwise and cumulative distribution $\mathbf{F_w(x) = \min\{1, wx\}}$. Estimators for PRI sketches [26] have (nearly) minimum sum of per-key variances $\sum_{i \in I} \text{VAR}(a(i))$ [53].

**Adjusted weights.** As mentioned in the introduction one technique to obtain estimators for the weights of keys is by assigning an adjusted weight $a(i) \geq 0$ to each key $i$ in the sample (adjusted weight $a(i) = 0$ is implicitly assigned to keys not in the sample). The adjusted weights are assigned such that $E[a(i)] = w(i)$, where the expectation is over the randomized algorithm choosing the sample. Using adjusted weights we can estimate the weight of any subpopulation $J \subset I$ by $\sum_{j \in J} a(j) = \sum_{j \in J | a(j) > 0} a(j)$. The estimate is easily computed from the sample assuming we have sufficient auxiliary information to tell for each key in the sample whether it belongs to $J$ or not. Moreover, for any numeric function $h()$ over keys' attributes such that $h(i) > 0$ only if $w(i) > 0$ and any subpopulation $J$, $\sum_{j \in J | a(j) > 0} a(j)h(j)/w(j)$ is an unbiased estimate of $\sum_{j \in J} h(j)$.

**Horvitz-Thompson** (HT). Let $\Omega$ be the distribution over sketches. If we know $p^{(\Omega)}(i) = \Pr\{i \in s | s \in \Omega\}$ for every $i \in s$ then we can assign to $i \in s$ the adjusted weight

$$a(i) = \frac{w(i)}{p^{(\Omega)}(i)}.$$

Since $a(i)$ is 0 when $i \notin s$, it is easy to see that $E[a(i)] = w(i)$. The estimator based on these adjusted weights is called the Horvitz-Thompson (HT) estimator [35]. It is well known and easy to see that these adjusted weights are unbiased and have minimal variance *for each key* for the particular distribution $\Omega$ over rank assignments.

HT **on a partitioned sample space** (HTP). This is a method to derive adjusted weights when we cannot determine $\Pr\{i \in s | s \in \Omega\}$ from the information contained in the sketch $s$ alone. For example, if $s$ is a bottom-$k$ sketch of $(I, w)$, then $\Pr\{i \in s | s \in \Omega\}$ generally depends on all the weights $w(i)$ for $i \in I$ and therefore cannot be determined from $s$.

For each key $i$ we consider a partition of $\Omega$ into equivalence classes. For a sketch $s$, let $P^i(s) \subset \Omega$ be the equivalence class of $s$. This partition must satisfy the following requirement: Given $s$ such that $i \in s$, we can compute the conditional probability $p^i(s) = \Pr\{i \in s' \mid s' \in P^i(s)\}$ from the information included in $s$.

We can therefore compute for all $i \in s$ the assignment $a(i) = w(i)/p^i(s)$ (implicitly, $a(i) = 0$ for $i \notin s$.) It is easy to see that within each equivalence class, $E[a(i)] = w(i)$. Therefore, also over $\Omega$ we have $E[a(i)] = w(i)$.

The variance of the adjusted weight $a(i)$ obtained using HTP depends on the particular partition in the following way. (This follows from the convexity of the variance.)

LEMMA 2.1. *[19] Consider two partitions of the sample space, such that one partition is a refinement of the other. Then the variance of $a(i)$ using* HTP *with the coarser partition is at most that of the* HTP *with the finer partition.*

**Rank Conditioning** (RC) **adjusted weights.** This is an HTP estimator for a single bottom-$k$ sketch [19]. The partition of $\Omega$ which we use for assigning an adjusted weight to $i$ is based on *rank conditioning*: For each possible rank value $\tau$ we have an equivalence class $P^i_\tau$ containing all sketches in which the $k$th smallest rank value assigned to a key other than $i$ is $\tau$. Note that if $i \in s$ then this is the $(k+1)$st smallest rank which is included in the sketch. It is easy to see that the inclusion probability of $i$ in a sketch in $P^i_\tau$ is $p^i_\tau = \mathbf{F}_{w(i)}(\tau)$.

Assume $s$ contains $i_1, \ldots, i_k$ and the $(k+1)$st smallest rank value $r_{k+1}$. Then for key $i_j$, we have $s \in P^{i_j}_{r_{k+1}}$ and $a(i_j) = \frac{w(i_j)}{\mathbf{F}_{w(i_j)}(r_{k+1})}$.

## 3. Combinations of bottom-k sketches

Consider a weighted set $I$, a set $\mathcal{S}$ of subsets of $I$, a family of rank functions $\mathbf{F}_w$ ($w > 0$), and a set of coordinated bottom-$k$ sketches $s_k(A, r)$ for $A \in \mathcal{S}$, where $r$ is drawn according to $\mathbf{F}_w$ ($w > 0$).

The *short combination of sketches* (SCS) of $\mathcal{S}$, denoted $\text{SCS}_k(\mathcal{S}, r)$, contains the prefixes of the sketches $s_k(A, r)$ ($A \in \mathcal{S}$) that include all keys with rank values smaller than $r_{k+1}(\mathcal{S}) = \min_{A \in \mathcal{S}} r_{k+1}(A)$. The SCS also includes the value $r_{k+1}(\mathcal{S})$. The SCS contains between $k$ and $|\mathcal{S}|k$ keys. Its size depends on the rank assignment. Its expected size is larger when sets are of similar weights and have fewer common keys.

The $\ell \geq k$ keys in the SCS are the $\ell$ least-ranked keys in the union $\bigcup_{A \in \mathcal{S}} A$ and $r_{k+1}(\mathcal{S}) = r_{\ell+1}(\bigcup_{A \in \mathcal{S}} A)$. Moreover, $\ell$ is *maximal* for which we can identify the $\ell$ least-ranked keys in the union from information available in the sketches of $\mathcal{S}$. For WS sketches, the SCS can be viewed as the outcome of weighted sampling without replacement (ppswor) from the union of the sets $\mathcal{S}$ until we obtain $k$ distinct samples from at least one of the sets in $\mathcal{S}$.

An important property of the SCS is that for every key $x$ in $\text{SCS}_k(\mathcal{S}, r)$ and a set $A \in \mathcal{S}$ we can determine if $x \in A$: Indeed $x \in A$ if and only if $x$ is in $s_k(A, r)$. The SCS is the maximal set of keys that are included in the union of the sketches and have this property.

The *long combination of sketches* (LCS) of $\mathcal{S}$, denoted $\text{LCS}_k(\mathcal{S}, r)$, includes all the information in the sketches $s_k(A, r)$, $A \in \mathcal{S}$.

The LCS includes the SCS, but we do not have complete set-membership information for all its keys. These definitions and relations are illustrated in Figure 1 through a detailed example of 4 sets defined over a ground set of 10 keys. The example demonstrates that the SCS and LCS contain more keys than the union-sketch.

In the sequel we derive estimators that reflect this relationship between combinations: SCS estimators are tighter than union-sketch estimators, reflecting the fact that the SCS contains the union-sketch. They are both applicable to arbitrary *select* predicates, reflecting the full membership information that is available for each included key. LCS based estimators are *tighter* than SCS based estimators, reflecting the fact that the LCS contains the SCS but LCS based estimators are *more limited* in that they are applicable only to restricted *select* predicates, reflecting the fact that we have less information for included keys.

## 4. Combination RC Estimators

We derive RC estimators for $\text{SCS}_k(\mathcal{S}, r)$ and $\text{LCS}_k(\mathcal{S}, r)$. Our RC estimators assign adjusted weights that are positive for all keys included in the respective combination (other keys are implicitly assigned adjusted weight of zero), are unbiased for all keys in $U = \bigcup_{A \in \mathcal{S}} A$, and have zero covariances.

Let $p(w, \tau) \equiv \lim_{x \to \tau^-} \mathbf{F}_w(x)$ be the probability than a key with weight $w$ obtains rank value that is smaller than $\tau$.

SCS RC **adjusted weights** $a^{(\text{SCS})}(i)$:

- $r_{k+1}(\mathcal{S}) \leftarrow \min_{A \in \mathcal{S}} r_{k+1}(A)$.
- $\text{SCS}_k(\mathcal{S}, r) \leftarrow \{i \in \bigcup_{A \in \mathcal{S}} s_k(A, r) \mid r(i) < r_{k+1}(\mathcal{S})\}$
- for all $i \in \text{SCS}_k(\mathcal{S}, r)$, assigned the adjusted weight

$$a^{(\text{SCS})}(i) \leftarrow \frac{w(i)}{p(w(i), r_{k+1}(\mathcal{S}))}. \quad (1)$$

(For WS sketches, $a^{(\text{SCS})}(i) = w(i)/(1 - \exp(-w(i)r_{k+1}(\mathcal{S})))$, and for PRI sketches $a^{(\text{SCS})}(i) = \max\{w(i), 1/r_{k+1}(\mathcal{S})\}$). Figure 2 demonstrates the computation of SCS RC adjusted weights and ( RC adjusted weights for the union sketch.

- **union-sketch** RC adjusted weights for $i_j \in s_3(\bigcup_{A \in \mathcal{S}} A, r)$:
$\tau = r_4(\bigcup_{A \in \mathcal{S}} A), a^{(\text{union})}(i_j) = w_j/p(w_j, \tau)$

| $\mathcal{S}$ | $\tau$ | $p(w, \tau)$ | $a^{(\text{union})}(i_j)$ |
|---|---|---|---|
| $A_1, A_2$ | 0.341 | $\min\{0.341w, 1\}$ | $\max\{w_j, 2.93\}$ |
| $A_1, A_2, A_3, A_4$ | 0.3 | $\min\{0.3w, 1\}$ | $\max\{w_j, 3.33\}$ |

- SCS RC **adjusted weights for** $i_j \in \text{SCS}_3(\mathcal{S}, r)$:
$\tau = r_4(\mathcal{S}), a^{(\text{SCS})}(i_j) = w_j/p(w_j, \tau)$

| $\mathcal{S}$ | $r_4(\mathcal{S})$ | $p(w, r_4(\mathcal{S}))$ | $a^{(\text{SCS})}(i_j)$ |
|---|---|---|---|
| $A_1, A_2$ | 0.73 | $\min\{0.73w, 1\}$ | $\max\{w_j, 1.37\}$ |
| $A_1, A_2, A_3, A_4$ | 0.599 | $\min\{0.599w, 1\}$ | $\max\{w_j, 1.67\}$ |

- LCS RC **adjusted weights for** $i_j \in \text{LCS}_3(\mathcal{S}, r), \mathcal{S} = \{A_1, A_2, A_3, A_4\}$.
Sets sorted by increasing $r_4(A_i)$: $A_3, A_4, A_1, A_2$

| $i_j$ | $i_7$ | $i_4$ | $i_2$ | $i_3$ | $i_{10}$ | $i_1$ | $i_6$ |
|---|---|---|---|---|---|---|---|
| $f(\mathcal{S}, r, i_j)$ | 1 | 4 | 2 | 1 | 2 | 1 | 2 |
| $\tau(\mathcal{S}, r, i_j)$ | 0.73 | 0.599 | 0.73 | 0.73 | 0.73 | 0.73 | 0.73 |
| $a^{(\text{LCS})}(i_j)$ | 1.37 | 3 | 2 | 1.37 | 1.37 | 1.37 | 1.37 |

Union-sketch, SCS/LCS RC adjusted weights for $\mathcal{S} = \{A_1, A_2\}$:

| key | $i_1$ | $i_2$ | $i_3$ | $i_4$ | $i_5$ | $i_6$ | $i_7$ | $i_8$ | $i_9$ | $i_{10}$ |
|---|---|---|---|---|---|---|---|---|---|---|
| $w_j$ | 1 | 2 | 1 | 3 | 1 | 1 | 1 | 1 | 1 | 1 |
| union | 0 | 2.93 | 2.93 | 0 | 0 | 0 | 2.93 | 0 | 0 | 0 |
| SCS/LCS | 1.37 | 2 | 1.37 | 0 | 0 | 1.37 | 1.37 | 0 | 0 | 1.37 |

Union-sketch, SCS, and LCS RC adjusted weights for $\mathcal{S} = \{A_1, A_2, A_3, A_4\}$:

| key | $i_1$ | $i_2$ | $i_3$ | $i_4$ | $i_5$ | $i_6$ | $i_7$ | $i_8$ | $i_9$ | $i_{10}$ |
|---|---|---|---|---|---|---|---|---|---|---|
| $w_j$ | 1 | 2 | 1 | 3 | 1 | 1 | 1 | 1 | 1 | 1 |
| union | 0 | 3.33 | 0 | 3.33 | 0 | 0 | 3.33 | 0 | 0 | 0 |
| SCS | 1.67 | 2 | 1.67 | 3 | 0 | 0 | 1.67 | 0 | 0 | 1.67 |
| LCS | 1.37 | 2 | 1.37 | 3 | 0 | 1.37 | 1.37 | 0 | 0 | 1.37 |

**Figure 2.** Upper box: Adjusted weights computation for example in Figure 1. SCS and LCS-adjusted weights for $\mathcal{S} = \{A_1, A_2\}$ are equal since $r_4(A_1) = r_4(A_2) = r_4(\{A_1, A_2\})$. Lower two tables: RC adjusted weights computed using the union-sketch, the SCS and the LCS.

We show that $a^{(\text{SCS})}$ are unbiased:

LEMMA 4.1. *For all* $i \in U$, $\mathrm{E}[a^{(\text{SCS})}(i)] = w(i)$.

PROOF. We apply HTP. For a key $i$ we partition the space of all rank assignments according to the rank values assigned to the keys $U \setminus \{i\}$. Consider a subspace $R$ in this partition. Fix some $r \in R$ and let

$$\tau(R) = \min\{ \begin{array}{l} \min\{r_k(A \setminus \{i\}) \mid A \in \mathcal{S}, i \in A\} \\ \min\{r_{k+1}(A) \mid A \in \mathcal{S}, i \notin A\} \end{array} \}.$$

Clearly $\tau(R)$ is independent of the choice of $r \in R$.

For $r \in R$, the key $i$ is included in $\text{SCS}_k(\mathcal{S}, r)$ if and only if $r(i) < \tau(R)$, which happens with probability $p(w(i), \tau(R))$. If indeed $i$ is included in $\text{SCS}_k(\mathcal{S}, r)$ then $r_{k+1}(\mathcal{S}) = \tau(R)$. □

We show that $a^{(\text{SCS})}$ have zero covariances:

LEMMA 4.2. *For* $i, j \in U, i \neq j$, $\text{COV}[a^{(\text{SCS})}(i), a^{(\text{SCS})}(j)] = 0$.

PROOF. We partition the space of rank assignments and show that in each set of the partition, $\mathrm{E}[a(i)a(j)] = w(i)w(j)$. The partition is according to the rank values assigned to all keys in $U \setminus \{i, j\}$. Let $R$ be a subspace in the partition, and let $r$ be a rank assignment in $R$. Define

$$\tau(R) = \min\{ \begin{array}{l} \min\{r_{k-1}(A \setminus \{i, j\}) \mid A \in \mathcal{S}, i, j \in A\}, \\ \min\{r_k(A \setminus \{i\}) \mid A \in \mathcal{S}, i \in A, j \notin A\}, \\ \min\{r_k(A \setminus \{j\}) \mid A \in \mathcal{S}, j \in A, i \notin A\}, \\ \min\{r_{k+1}(A) \mid A \in \mathcal{S}, i, j \notin A\} \end{array} \}.$$

- **Keys:**

| key | $i_1$ | $i_2$ | $i_3$ | $i_4$ | $i_5$ | $i_6$ | $i_7$ | $i_8$ | $i_9$ | $i_{10}$ |
|---|---|---|---|---|---|---|---|---|---|---|
| $w_j$ | 1 | 2 | 1 | 3 | 1 | 1 | 1 | 1 | 1 | 1 |
| $u_j$ | 0.487 | 0.72 | 0.3 | 0.832 | 0.765 | 0.599 | 0.131 | 0.886 | 0.73 | 0.341 |
| $r_j = \frac{u_j}{w_j}$ | 0.487 | 0.36 | 0.3 | 0.208 | 0.765 | 0.599 | 0.131 | 0.886 | 0.73 | 0.341 |

- **Sets:**

$A_1 = \{i_1, i_3, i_5, i_7, i_9\}$
$A_2 = \{i_1, i_2, i_5, i_6, i_9, i_{10}\}$
$A_3 = \{i_3, i_4, i_5, i_6, i_7\}$
$A_4 = \{i_2, i_4, i_6, i_8, i_{10}\}$

- Keys sorted by increasing ranks (with matrix showing set memeberships of all keys):

| keys | $i_7$ (0.131) | $i_4$ (0.208) | $i_2$ (0.36) | $i_3$ (0.3) | $i_{10}$ (0.341) | $i_1$ (0.487) | $i_6$ (0.599) | $i_9$ (0.73) | $i_5$ (0.765) | $i_8$ (0.886) |
|---|---|---|---|---|---|---|---|---|---|---|
| $A_1$ | √ | × | × | √ | × | √ | × | √ | √ | × |
| $A_2$ | × | × | √ | × | √ | × | √ | √ | √ | × |
| $A_3$ | √ | √ | × | √ | × | × | √ | × | √ | × |
| $A_4$ | × | √ | √ | × | √ | × | √ | × | × | √ |

- Table showing bottom-3 sketches $s_3(A_i, r)$, (3 least-ranked keys of $A_i$ and $r_4(A_i)$, the $4^{\text{th}}$-smallest rank value):

| $A$ | $s_3(A, r)$ | $r_4(A)$ |
|---|---|---|
| $A_1$ | $i_7$ (0.131), $i_3$ (0.3), $i_1$ (0.487) | 0.73 |
| $A_2$ | $i_2$ (0.36), $i_{10}$ (0.341), $i_6$ (0.599) | 0.73 |
| $A_3$ | $i_7$ (0.131), $i_4$ (0.208), $i_3$ (0.3) | 0.599 |
| $A_4$ | $i_4$ (0.208), $i_2$ (0.36), $i_{10}$ (0.341) | 0.599 |

- For $\mathcal{S} = \{A_1, A_2\}$ and $\mathcal{S} = \{A_1, A_2, A_3, A_4\}$, keys included in:
  ⋄ $s_3(\bigcup_{A \in \mathcal{S}} A, r)$ (The union-sketch of $\mathcal{S}$)
  ⋄ $\text{SCS}_3(\mathcal{S}, r)$ (contains all keys in $\bigcup_{A_i \in \mathcal{S}} s_3(A_i, r)$ that have rank value below $r_4(\mathcal{S}) = \min_{A_i \in \mathcal{S}} r_4(A_i)$), and
  ⋄ $\text{LCS}_3(\mathcal{S})$ (contains all keys in $\bigcup_{A_i \in \mathcal{S}} s_3(A_i, r)$).

| S | combination type | content | # keys |
|---|---|---|---|
| $A_1, A_2$ | $s_3 (\bigcup_{A \in \mathcal{S}} A, r)$ | $i_7$ (0.131), $i_2$ (0.36), $i_3$ (0.3) | 3 |
| $A_1, A_2$ | $\text{SCS}_3 (\mathcal{S}, r)$ | $i_7$ (0.131), $i_2$ (0.36), $i_3$ (0.3), $i_{10}$ (0.341), $i_1$ (0.487), $i_6$ (0.599) | 6 |
| $A_1, A_2$ | $\text{LCS}_3 (\mathcal{S}, r)$ | $i_7$ (0.131), $i_2$ (0.36), $i_3$ (0.3), $i_{10}$ (0.341), $i_1$ (0.487), $i_6$ (0.599) | 6 |
| $A_1, A_2, A_3, A_4$ | $s_3 (\bigcup_{A \in \mathcal{S}} A, r)$ | $i_7$ (0.131), $i_4$ (0.208), $i_2$ (0.36) | 3 |
| $A_1, A_2, A_3, A_4$ | $\text{SCS}_3 (\mathcal{S}, r)$ | $i_7$ (0.131), $i_4$ (0.208), $i_2$ (0.36), $i_3$ (0.3), $i_{10}$ (0.341), $i_1$ (0.487) | 6 |
| $A_1, A_2, A_3, A_4$ | $\text{LCS}_3 (\mathcal{S}, r)$ | $i_7$ (0.131), $i_4$ (0.208), $i_2$ (0.36), $i_3$ (0.3), $i_{10}$ (0.341), $i_1$ (0.487), $i_6$ (0.599) | 7 |

**Figure 1.** Example shows a set $I$ of 10 keys $i_1, \ldots, i_{10}$ with respective weights $w_1, \ldots, w_{10}$ and 4 subsets $A_1, \ldots, A_4$; a random rank assignment $r$ for $I$, using priority ranks (for each key $i_j$, draw $u_j \in U[0, 1]$ and compute rank value $r_j = u_j/w_j$); bottom-3 sketches $s_3(A_i, r)$ for $i = 1, \ldots, 4$; for $\mathcal{S} = \{A_1, A_2\}$ and $\mathcal{S} = \{A_1, A_2, A_3, A_4\}$, keys included in the union-sketch, the SCS, and the LCS of S.

Clearly $\tau(R)$ is independent of the choice of $r \in R$. For $r \in R$, it is easy to see that $i$ and $j$ are both included in the SCS if and only if $r(i) < \tau(R)$ and $r(j) < \tau(R)$, which happens with probability $p(w(i), \tau(R))p(w(j), \tau(R))$. Otherwise either $i$ or $j$ is not included in the SCS and $a(i)a(j) = 0$. In the case they are both included, $r_{k+1}(\mathcal{S}) = \tau(R)$, and therefore they are assigned adjusted weights of $w(i)/p(w(i), \tau(R))$ and $w(j)/p(w(j), \tau(R))$, respectively. It follows that

$$\begin{aligned} \mathrm{E}[a(i)a(j)] &= \frac{p(w(i), \tau(R))p(w(j), \tau(R))w(i)w(j)}{p(w(i), \tau(R))p(w(j), \tau(R))} \\ &= w(i)w(j) \,. \end{aligned}$$

□

LCS RC **adjusted weights** $a^{(\text{LCS})}(i)$ :

- Sort the sets $A \in \mathcal{S}$ by increasing $r_{k+1}(A)$ into the ordered set $A_1, A_2, \ldots, A_{|\mathcal{S}|}$ ($r_{k+1}(A_i) \leq r_{k+1}(A_j)$ if $i < j$).
- For all $i \in \text{LCS}_k(\mathcal{S}, r)$:

$$f(\mathcal{S}, r, i) \leftarrow \arg\max_h i \in s_k(A_h, r)$$
$$\tau(\mathcal{S}, r, i) \leftarrow r_{k+1}(A_{f(\mathcal{S}, r, i)}).$$
$$a^{(\text{LCS})}(i) \leftarrow \frac{w(i)}{p(w(i), \tau(\mathcal{S}, r, i))} \,. \quad (2)$$

Figure 2 demonstrates the computation of RC LCS adjusted weights.

LEMMA 4.3. *For all $i \in U$, $\mathrm{E}[a^{(\text{LCS})}(i)] = w(i)$.*

PROOF. For a key $i \in U$, we partition the space of all rank assignments according to the rank values of keys in $U \setminus \{i\}$. Consider a subspace $R$ in this partition, let $r$ be a rank assignment in $R$, and let $\tau(R) = \max_{A \in \mathcal{S} | i \in A} r_k(A \setminus \{i\})$, which is independent of the choice of $r \in R$.

For $r \in R$, the key $i$ is included in $\text{LCS}_k(\mathcal{S}, r)$ if and only if $r(i) < \tau(R)$. This happens with probability $p(w(i), \tau(R))$ and when it happens we clearly have that $r_{k+1}(A_{i_{f(\mathcal{S}, r, i)}}) = \tau(R)$, which implies the lemma. □

| | condition | relevant sets $\mathcal{S}$ | keys | weight | RC union | RC SCS | RC LCS | best comb |
|---|---|---|---|---|---|---|---|---|
| $P_1$ | $i_j \in \bigcup_{i \in [2]} A_i \wedge (j < 8 \vee j \geq 4)$ | $A_1, A_2$ | $i_5, i_6, i_7$ | 3 | 2.93 | 2.74 | 2.74 | LCS$_3$ |
| $P_2$ | $i_j \in \bigcap_{i \in [2]} A_i \wedge (j < 8 \vee j \geq 4)$ | $A_1, A_2$ | $i_5$ | 1 | 0 | 1.37 | – – – | SCS$_3$ |
| $P_3$ | $i_j \in$ at least two out of $A_1, \ldots, A_4$ | $A_1, A_2, A_3, A_4$ | $i_1, \ldots, i_7, i_9, i_{10}$ | 12 | 10 | 11.68 | – – | SCS$_3$ |
| $P_4$ | $(i_j \in \bigcup_{i \in [4]} A_i \wedge j$ is odd | $A_1, A_2, A_3, A_4$ | $i_1, i_3, i_5, i_7, i_9$ | 5 | 3.33 | 5 | 4.1 | LCS$_3$ |

**Figure 3.** Example predicates for the dataset in Figure 1. Table shows for each predicate $P$, a minimum set of relevant sets, all keys that satisfy $P(i)$, weight of these keys, best applicable combination, and RC union, RC SCS, and RC LCS estimates, based on adjusted weights computation in Figure 2. (LCS adjusted weight is not shown for predicates where LCS is not applicable).

LEMMA 4.4. *For $i, j \in U$, $i \neq j$, $\text{COV}[a^{(\text{LCS})}(i), a^{(\text{LCS})}(j)] = 0$.*

PROOF. Consider the subspace where all ranks of keys other than $i$ and $j$ are fixed. We compute $E[a(i)a(j)]$ in this subspace.

Let $\mathcal{S}_i$ be the collection of sets in $\mathcal{S}$ that contain key $i$ and do not contain key $j$. Let $\mathcal{S}_j$ be the collection of sets in $\mathcal{S}$ that contain key $j$ and do not contain key $i$. Finally $\mathcal{S}_{i,j}$ be the collection of sets in $\mathcal{S}$ containing both $i$ and $j$.

Define $r^{-i} = \max(\{r_k(A \setminus \{i\}) \mid A \in \mathcal{S}_i\} \cup \{r_{k-1}(A \setminus \{i,j\}) \mid A \in \mathcal{S}_{i,j}\})$, $r^{-j} = \max(\{r_k(A \setminus \{j\}) \mid A \in \mathcal{S}_j\} \cup \{r_{k-1}(A\setminus\{i,j\}) \mid A \in \mathcal{S}_{i,j}\})$, and $r^{-i,j} = \max(\{r_k(A\setminus\{i,j\}) \mid A \in \mathcal{S}_{i,j}\}$.

We split into cases according to the relations between $r^{-i}$, $r^{-j}$, and $r^{-i,j}$. If $r^{-i,j} \leq \min\{r^{-i}, r^{-j}\}$ or if $\max\{r^{-i}, r^{-j}\} \leq r^{-i,j}$, then $i$ and $j$ are both included (and $a(i)a(j) > 0$) if and only if $r(i) < r^{-i}$ and $r(j) < r^{-j}$. In which case $a(i) = \frac{w(i)}{p(w(i), r^{-i})}$ and $a(j) = \frac{w(j)}{p(w(j), r^{-j})}$. Therefore, under this conditioning,

$$E[a(i)a(j)] = p(w(i), r^{-i})p(w(j), r^{-j}) \frac{w(i)}{p(w(i), r^{-i})} \frac{w(j)}{p(w(j), r^{-j})}$$
$$= w(i)w(j).$$

The remaining case is $r^{-i} < r^{-i,j} < r^{-j}$ (the case $r^{-j} < r^{-i,j} < r^{-i}$ is symmetric). $j$ is included if and only if $r(j) \leq r^{-j}$, in which case $a(j) = \frac{w(j)}{p(w(j), r^{-j})}$. The inclusion condition and adjusted weight of $i$ if included depend on $r(j)$, but if we fix $r(j)$, from the proof of Lemma 4.3, $E[a(i)] = w(i)$. That is, if $a(i|y, x)$ denotes the adjusted weight of $i$ if $r(j) = x$ and $r(i) = y$, then for all $x$, $\int_0^\infty a(i|y, x)dy = w(i)$. Therefore,

$$E[a(i)a(j)] = \int_0^{r^{-j}} \mathbf{f}_{w(j)}(x) \frac{w(j)}{p(w(j), r^{-j})} \int_0^\infty a(i|y, x)dy dx$$
$$= \int_0^{r^{-j}} \mathbf{f}_{w(j)}(x)dx \frac{w(j)}{p(w(j), r^{-j})} w(i)$$
$$= p(w(j), r^{-j}) \frac{w(j)}{p(w(j), r^{-j})} w(i) = w(j)w(i)$$

□

Consider the set $\mathcal{S}$ of subsets of $I$, a family of rank functions, and coordinated bottom-$k$ sketches $s_k(A, r)$ for $A \in \mathcal{S}$. We compare the three RC adjusted weight assignments $a^{(C)}(i)$ ($i \in U$), where $C$ is

- union: single-sketch RC adjusted weights on the sketch of the union $s_k(U, r)$
- SCS: SCS RC adjusted weights on $\text{SCS}_k(\mathcal{S}, r)$
- LCS: LCS RC adjusted weights on $\text{LCS}_k(\mathcal{S}, r)$

LEMMA 4.5. *For any $J \subset U$,*

$$\text{VAR}[a^{(\text{LCS})}(J)] \leq \text{VAR}[a^{(\text{SCS})}(J)] \leq \text{VAR}[a^{(union)}(J)].$$

PROOF. Because all methods have zero covariances between different keys, it suffices to establish that relation for the variances of per-key adjusted weights, that is, for any $i \in U$, $\text{VAR}[a^{(\text{LCS})}(i)] \leq \text{VAR}[a^{(\text{SCS})}(i)] \leq \text{VAR}[a^{(union)}(i)]$.

Consider a key $i$ and a subspace $R$ of the sample space of rank assignments such that the rank values of all other keys are fixed. It suffices to show the variance relation in each such subspace.

Let $q^{(union)}(R, i)$, $q^{(\text{SCS})}(R, i)$, $q^{(\text{LCS})}(R, i)$ be the probabilities conditioned on $R$ that $i$ is included in the respective combination. Since the probability $p(w(i), \tau)$ is decreasing with $\tau$ and $r_{k+1}(\bigcup_{A \in \mathcal{S}} A) \leq r_{k+1}(\mathcal{S}) \leq \tau(\mathcal{S}, r, i)$, we have that $q^{(union)}(R, i) \leq q^{(\text{SCS})}(R, i) \leq q^{(\text{LCS})}(R, i)$.

For any combination $C \in \{union, \text{LCS}, \text{SCS}\}$ the adjusted weight in $R$ is the HT estimator $a^{(C)}(i) = w(i)/q^{(C)}(R, i)$. The variance of $a^{(C)}(i)$ is decreasing with the probability $q^{(C)}(R, i)$, which concludes the proof. □

## 5. Computing Estimates

The input to our estimation procedure is a set of coordinated bottom-$k$ sketches $s_k(A, r)$ for sets $A \subset I$, $A \in \mathcal{A}$, and a weight query specified by a predicate $P : I$. The desired output is an estimate of $\sum_{i \in I | P(i)} w(i)$.

We use the following two definitions:

- A set of *relevant sets* $\mathcal{S} \subset \mathcal{A}$ for a predicate $P$ is a set of sets that suffices to determine the keys that satisfy $P$. For example, for the query "term is present in at least 2 out of books $A$, $B$, $C$," the relevant sets are $A$, $B$, and $C$. The query "term present in $A$ and not in $C$" has relevant sets $A$ and $C$. In both cases, these are minimum relevant sets. The first step of processing the query for $P$ is determining (preferably a minimal) set $\mathcal{S}$ of relevant sets.

- The *best applicable combination* for $P$ is the most inclusive combination $C \in \{\text{SCS}, \text{LCS}\}$ that allows us to evaluate $\sum_{i \in C | P(i)} a(i)$ using information that is available in the sketches of $\mathcal{S}$. Since we get better estimates with the LCS, we should use the LCS when it is applicable.

We can evaluate $P(i)$ for all $i \in \text{SCS}$ for general $P$. This is because we have full membership information in $\mathcal{S}$ sets for all keys in $\text{SCS}(\mathcal{S})$. For $i \in \text{LCS}$, we can determine membership of $i$ only in sets $A \in \mathcal{S}$ such that $r(i) \leq r_{k+1}(A)$. Since the combination must be applicable to all rank assignments, we can apply the LCS only to predicates $P$ that have the form of an attribute-based condition over keys in $\bigcup_{A \in \mathcal{S}} A$. As an example, the SCS is the best

applicable combination for the intersection of two sets $A \cap B$.[2]

> **Input:** set of coordinated bottom-$k$ sketches $s_k(A, r)$ for sets $A \in \mathcal{A}$; predicate $P$
> - Analyze $P$ to determine:
>   ⋄ A (minimum) set $\mathcal{S}$ of "relevant sets."
>   ⋄ The best applicable combination $C \in \{\text{SCS}, \text{LCS}\}$:
>     If $P$ is an attribute-based condition over $\bigcup_{A \in \mathcal{S}} A$, $C \leftarrow \text{LCS}$.
>     Else, $C \leftarrow \text{SCS}$.
> - Retrieve the sketches $s_k(A, r)$ of the sets $A \in \mathcal{S}$.
> - Compute adjusted weights $a^{(C)}(i)$ for $i \in C_k(\mathcal{S}, r)$ using (1), if $C \equiv \text{SCS}$, or (2), if $C \equiv \text{LCS}$.
> - **Output:** $\sum_{i \in C | P(i)} a(i)$.

Note that once adjusted weights are computed, they can be applied to multiple predicates that share the same relevant set $\mathcal{S}$ and best applicable combination $C$.

Figure 3 illustrates the evaluation of an approximate weight for some example predicates.

## 6. Poisson sampling

We elaborate on the relation of bottom-$k$ (order) sampling and Poisson sampling. In particular, we discuss coordinated sketches based on Poisson samples, estimators applicable to these sketches and their relation to our estimators for order samples, and computation issues.

Poisson sampling is a classic sampling method where each key has an independent inclusion probability [33] which depends on the weight of the key. Order (bottom-$k$) sampling [46, 47, 44, 45, 47] was initially developed as a twist on Poisson sampling intended to achieve fixed-size samples. Literature in the computer science field (re-)introduced order sampling as an alternative to $k$-mins sampling and as a weighted reservoir sampling scheme [12, 7, 42, 51, 17, 27, 26, 2, 32, 19].

Following [33, 47] but using our terminology, a Poisson sample of a weighted set $(I, w)$ with respect to a family of distribution functions $\mathbf{f}_w$ ($w > 0$) and a value $\tau$ is obtained by drawing a random rank assignment on $(I, w)$ and including all keys with rank value $r(i) < \tau$. (Recall that an order sample with respect to the same rank assignment and a fixed $k$ includes the $k$ keys with smallest rank values.)

The probability that a key $i$ is included in the sample is $p(w(i), \tau)$ and inclusion probabilities of different keys are independent. (Recall that inclusion probabilities are dependent with bottom-$k$ sampling).

There is a natural correspondence between Poisson and order samples [44, 45, 47]: For a Poisson sample with a given $\tau$, the corresponding order sample has size $k = \sum_i p(w(i), \tau)$ equal to the *expected* sample size of the Poisson sample. This correspondence facilitates the comparison of estimators over the two sampling methods.

IPPS (inclusion probability proportion to size) sampling [33, 49] includes each key with probability proportional to its weight. These inclusion probabilities are known to minimize variance for a given sample size. The order sampling equivalent of IPPS is priority sampling (previously introduced as Sequential Poisson Sampling) [44, 45, 47, 26]. Szegedy [53] recently established that the estimator of [26] for priority sample of size $k + 1$ has a sum of per-key variances that is at most that of an IPPS Poisson sample with $\tau$ that corresponds to $k$.

Adjusted weights for Poisson sampling, $w(i)/p(w(i), \tau)$ for a key $i$, are a straightforward application of the HT estimator. In contrast, adjusted weights for order samples were only recently [26, 19] derived. Szegedy's result [53] means that we can simultaneously enjoy the fixed sample size of order sampling and (nearly) optimal variance of IPPS Poisson sampling.

Poisson sampling can be performed on a data stream in a scalable way only if IPPS sampling is used. Indeed, Poisson sampling with respect to a *fixed* $\tau$, can be computed in a straightoffward way in a single pass that indepedently samples each key. Typically, however, resource constraints limit sample size. For Poisson sampling, this means that we want to set $\tau$ so that the *expected* sample size is $k$, that is, $\tau$ that solves the equation $k = \sum_i p(w(i), \tau)$. In a reservoir or data stream setting, we need to track the solution of $k = \sum_i p(w(i), \tau)$ with respect to the prefix of the keys seen so far, which is possible to do efficiently for IPPS sampling (and uniform weighted keys as a special case). Generally, fixed expected-size sampling requires two passes over the data set.

**Coordinated Poisson samples and estimators.** Coordinated Poisson samples for $\mathcal{S}$ are such that there is a different (fixed) $\tau_A$ for each set $A \in \mathcal{S}$ and the Poisson sample of $A$ includes all keys with $r(i) < \tau_A$. We outline SCS-like and LCS-like adjusted weights over coordinated Poisson samples. The expressions resemble, but are simpler, than the ones we present for order samples (Section 4). We highlight properties of these estimators. The analysis (just like for s Poisson sample of a single set) is straightforward.

To express SCS-*like adjusted weights* for Poisson samples, define $\tau_\mathcal{S} = \min_{A \in \mathcal{S}} \tau_A$. The set of keys in the union of the Poisson samples of $A \in \mathcal{S}$ that have $r(i) < \tau_\mathcal{S}$ constitute a Poisson sample with $\tau_\mathcal{S}$ of the union of $\mathcal{S}$. For each key in this sample, we know which sets $A \in \mathcal{S}$ it is a member of. We can therefore use the adjusted weights $w(i)/p(w(i), \tau)$ for these keys and obtain unbiased estimator for any selection predicate. This derivation generalizes the estimator used in [30, 31] for keys with uniform weights.

LCS-*like adjusted weights* are positive for all keys included in the union of the Poisson samples of $\mathcal{S}$. The adjusted weight of a key $i$ is computed as follows. Let $A$ be the set with largest $\tau_A$ such that $i$ is included in the sample of $A$. The adjusted weight is then $w(i)/p(w(i), \tau_A)$.

Zero covariances of the SCS-like and LCS-like adjusted weights of different keys are immediate from independence. Just like with order sampling, LCS-like adjusted weights dominates SCS-like adjusted weights (have at most the variance on all subpopulations) but the LCS estimator is applicable only to selection predicates that are attribute based selections from the union of $\mathcal{S}$. Coordinated Poisson and coordinated order samples can be compared if we set the Poisson sampling $\tau$ value of each set to correspond to expected sample size of $k$.

**Comparing estimators.** Empirical evaluation of combination estimators indicates similar performance. We suspect that Szegedy's [53] result on the relation of Priority (order) sampling and threshold (Poisson) sampling generalizes to the respective SCS and LCS variants.

**Scalability of sampling.** Recall that coordinated bottom-$k$ samples can be computed efficiently over explicit or implicit representation of the data sets. Coordinated Poisson samples over explicitly represented data sets can be computed efficiently in a single pass for IPPS sampling or if fixed $\tau$ values are used for each set. It seems that (even for IPPS or uniform weights), there may not be a scalable method for computing Poisson samples with fixed expected sample size or all-distances sketches over implicitly represented sets. Intuitively, the difficulty is that respective exact $\tau$ values must be determined for all sets. With uniform weights, exact $\tau$ values correspond to exact sizes of the sets. But it seems that determining exact sizes (of say, all reachability sets in a graph) is considerably harder than the respective estimation problem [12].

---
[2] We can still use the LCS indirectly to estimate $w(A \cap B)$ using the inclusion-exclusion formula $w(A \cap B) = w(A) + w(B) - w(A \cup B)$. But this estimator does not perform well (see Section 9).

Our combination estimators for bottom-$k$ samples offer a "best of all worlds solution:" They match the performance benefits of these Poisson multiple-set estimators and have the more desirable framework of order sampling (fixed sample size and scalable computation in more applications.)

## 7. Unbiased selectivity estimators

We estimate selectivities through *adjusted selectivities* $\rho(i)$ such that $\mathrm{E}[\rho(i)] = w(i)/w(U)$ (for all $i \in U$).

We consider three types of sketches $M \in \{\text{WSR}, \text{WSRD}, \text{WSRC}\}$ based on sampling with replacement from $U$. For an infinite sequence $s$ of weighted sampling with replacement from $U$, we consider sampling with the following stopping rules. (i) WSR ($k$-mins): after $k$ (not necessarily distinct) samples, (ii) WSRD: when seeing the $k+1$ distinct key, (iii) WSRC: with respect to $\mathcal{S}$, when, for at least one set $A \in \mathcal{S}$, we see the $(k+1)$st distinct key from $A$.

The respective $M$ sketch is a set of keys and multiplicities $c^{(M)}(i,s)$ ($i \in U$), (the number of times $i$ was sampled before stopping). $c^{(M)}(U,s)$ denotes the sum of multiplicities of keys.

LEMMA 7.1. *For $M \in \{\text{WSR}, \text{WSRD}, \text{WSRC}\}$, $\rho_1^{(M)}(i,s) = c^{(M)}(i,s)/c^{(M)}(U,s)$ are correct adjusted selectivities.*

PROOF. WSR: By definition, $c^{(M)}(U,s) \equiv k$ and we obtain the WSR $k$-mins estimator in [12, 7]. This well-known estimator, used in [12, 7] to estimate the resemblance of $A_1$ and $A_2$ (the sum of multiplicities of keys from $A_1 \cap A_2$ in the WSR $k$-mins sketch of $A_1 \cup A_2$, divided by $k$.), assigns to each key an adjusted selectivity equals to its multiplicity in the sketch times $1/k$.

WSRD: Consider a key $i$. Partition the probability space so that in each set of the partition the number of samples of keys from $U \setminus \{i\}$ until we get $k$ distinct keys from $U \setminus \{i\}$ is fixed. We will show that $\rho(i)$ is an unbiased selectivity in each subspace. Consider a subspace where the number of samples of keys from $U \setminus \{i\}$ until we get $k$ distinct keys from $U \setminus \{i\}$ is $\ell$. (Notice that $\ell \geq k$.) The estimator $\rho(i)$ in this subspace is $\frac{c(i,s)}{c(i,s)+\ell-1}$. This is because if we do not sample $i$ by the time we get $k$ distinct keys from $U \setminus \{i\}$ then $c(i,s) = 0$ as well as $\rho(i)$, and otherwise $c(U,s) = c(i,s) + \ell - 1$ and therefore $\rho(i) = \frac{c(i,s)}{c(i,s)+\ell-1}$.

The number of times $i$ is sampled between two samples from $U \setminus \{i\}$ is geometrically distributed with parameter $p = w(i)/w(U)$. Therefore we need to show that

$$\sum_{i_1=0}^{\infty} \cdots \sum_{i_\ell=0}^{\infty} p^{\sum_{j=1}^{\ell} i_j} (1-p)^\ell \frac{\sum_{j=1}^{\ell} i_j}{\ell - 1 + \sum_{j=1}^{\ell} i_j} = p. \quad (3)$$

By combining together terms in which $\sum_{j=1}^{\ell} i_j = t$ in the left side of (3) we obtain that

$$\sum_{t=1}^{\infty} \binom{t+\ell-1}{\ell-1} p^t (1-p)^\ell \frac{t}{t+\ell-1}$$
$$= (1-p)^\ell \sum_{t=1}^{\infty} \binom{t+\ell-2}{\ell-1} p^t = (1-p)^\ell p \sum_{t=1}^{\infty} \binom{t+\ell-2}{\ell-1} p^{t-1}$$
$$= (1-p)^\ell p \sum_{t=0}^{\infty} \binom{t+\ell-1}{\ell-1} p^t = (1-p)^\ell p (1 + p + p^2 + \cdots)^\ell$$
$$= p$$

WSRC: For subsets $A \in \mathcal{S}$ such that $i \in A$ consider the occurrence of the $k$th distinct key from $A \setminus \{i\}$ and for subsets such that $i \notin A$ consider the occurrence of the $(k+1)$st distinct key from $A$. Fix $\ell$ and consider the subspace of the probability space where the total number of samples until and including the first among these occurrences. If $i$ is sampled at least once, then there are $\ell - 1$ samples from $U \setminus \{i\}$ in the WSRC sketch. The number of times $i$ is sampled between two samples from $U \setminus \{i\}$ is geometrically distributed with parameter $w(i)/w(U)$. The proof proceeds as the proof for WSRD sketches. □

LEMMA 7.2. *For $M \in \{\text{WSR}, \text{WSRD}, \text{WSRC}\}$:*
- *For $i \neq j \in U$, $\mathrm{COV}[\rho_1^{(M)}(i,s), \rho_1^{(M)}(j,s)] \leq 0$.*
- $\sum_{i \in U} \rho_1^{(M)}(i,s) = 1.$

LEMMA 7.3. *For $M \in \{\text{WSR}, \text{WSRD}, \text{WSRC}\}$ and $J \subset U$:*

$$\mathrm{VAR}[\rho_1^{(\text{WSRC})}(J,s)] \leq \mathrm{VAR}[\rho_1^{(\text{WSRC})}(J,s)] \leq \mathrm{VAR}[\rho_1^{(\text{WSR})}(J,s)].$$

### 7.1 Sampling without replacement

For a bottom-$k$ sketch when all keys have equal weights, Broder observed [6, 42] that the fraction of keys in the sketch of the union $A_1 \cup A_2$ that are contained in $A_1 \cap A_2$ is an unbiased estimator of the Jaccard coefficient. More generally, adjusted selectivities of $\rho^{(\text{WS})}(i) = 1/k$ are correct for WS sketches when the keys have equal weights.

This is not true anymore if keys have different weights as the following simple example shows. Consider two subsets $A_1 = \{i_1, i_2, i_3\}$ and $A_2 = \{i_1, i_4\}$. The union contains four keys $\{i_1, i_2, i_3, i_4\}$. Let the corresponding weights be $w(i_1) = 4$ and $w(i_j) = 1$ for $j > 1$. The intersection of the two sets is $\{i_1\}$. The resemblance is $4/7$. Consider a bottom-2 WS-sketch of $A_1 \cup A_2$. The probability that $i_1$ appears (first or second) in the sketch is $4/7 + (3/7) * (4/6) = 6/7$ in that case, the respective fraction is $4/5$ (since the other key in the sketch has weight 1). Otherwise, $i_1$ does not appear in the sketch and the fraction is zero. Therefore, the expectation of the fraction is $(6/7)(4/5) > 4/7$. If we use the fraction of coordinates of $A_1 \cap A_2$ in the sample instead of the fraction of weights, we obtain $3/7 < 4/7$.

Unbiased selectivity estimators for WS bottom-$k$ sketches can be obtained via a *mimicking process* [17]. The mimicking process is a randomized algorithm that inputs a WS sketch and output a sequence of "emulated" weighted samples with replacement. We can also apply mimicking to an SCS of WS bottom-$k$ sketches by arranging the keys in the SCS by increasing rank values and using this as an input to the process.

If we stop the process when $k$ keys (not necessarily distinct) are drawn, we obtain a WSR-sketch. If we continue until we see the $(k+1)$st distinct key, which exhausts the "information" in the

WS sketch, we obtain a WSRD sketch. If applied to an SCS until the information is exhausted, we obtain a WSRC *sketch*.

Mimicking allows us to carry over unbiased estimators applicable to WSR, WSRD, and WSRC sketches to WS sketches and SCS combinations.

**Tighter estimators.** The adjusted selectivities $\rho_1^{(M)}$ ($i \in s$) have the desirable qualities of (i) non-positive covariances between different keys and (ii) adjusted selectivities sum up to one. (See [19, 13, 54] for a discussion of these qualities.)

We obtain tighter estimators than $\rho_1^{(M)}$, that share these qualities but have a lower sum of per-key variances.

Mimicking is a random process and therefore, each WS sketch or SCS corresponds to a probability distribution $D$ over WSR and WSRD (for SCS also WSRC) sketches. Tighter estimators are obtained by taking the expectation of $\rho_1^{(M)}$ (or average over multiple draws) over $D$. We can get even tighter estimators by looking at the expectation of this estimator over equivalence classes of WS sketches (or SCS combinations). Equivalence class can include all sketches/combinations with same rank ordering of keys, obtained by redrawing the ranks of keys, or (if total weight is available) containing the same set of keys [17, 19]. One interesting corollary is the following:

LEMMA 7.4. *If all weights are equal, then $\rho^{SCS}(i) = 1/\ell$ for all $i \in \text{SCS}_k(\mathcal{S}, r)$, where $\ell = |\text{SCS}_k(\mathcal{S}, r)|$, are correct adjusted selectivities.*

PROOF. Redrawing the rank values of the first $\ell$ keys in $U$ does not change the SCS. The resulting distribution is symmetric for all $\ell$ keys and therefore the expectation of $\rho_1(i)$ is the same. □

The adjusted selectivities $\rho^{SCS}(i) = 1/\ell$ are superior to $\rho^{WS}(i) = 1/k$ (and in particular, improve over classic union-sketch resemblance estimator [6, 42]). Both estimators have symmetric non-negative covariances and the adjusted selectivities sum up to 1. However, $\text{VAR}[\rho^{WS}(i)] = N/k - 1 \geq \text{VAR}[\rho^{SCS}(i)] = N/k' - 1$, where $N$ is the total number of keys and $k' = 1/E[1/\ell]$, where $\ell \geq k$ is the number of keys in the SCS. (Let $p_\ell$ be the probability that the SCS contains $\ell$ keys. We have $\text{VAR}[\rho^{SCS}(i)] = \sum_\ell p_\ell(N/\ell - 1) = N/k' - 1$.) We typically have $k' \approx E[\ell]$. Section 9 includes an evaluation of this estimator relative to the classic union-sketch estimators for Jaccard coefficient.

## 7.2 Sampling with replacement

We derive tighter estimators than $\rho_1^{(M)}$ for WSR, WSRD, and WSRC sketches by considering the expectation of $\rho_1^{(M)}$ over equivalence classes that correspond to a partition of the sample space of sequences $s$. These estimators can be used with the mimicking process.

For each $s$ and $M \in \{\text{WSR}, \text{WSRD}, \text{WSRC}\}$, let $s^{(M)} = \{i | c^{(M)}(i, s) \geq 1\}$ be the set of distinct keys in the corresponding sketch. For each key $i \in s$, we know $w(i)$.[3]

**WSR:** Each equivalence class contains all $s$ such that the corresponding WSR sketches share the same set $s^{(WSR)}$ of distinct keys (the $k$ first samples have the keys $s^{(WSR)}$). For $s$, let $L(s)$ be the equivalence class containing $s$. (That is, $s' \in L(s)$ if and only if $c^{(WSR)}(i, s') \geq 1$ for all $i \in s^{(WSR)}$ and $c(i, s') = 0$ for $i \in U \setminus s^{(WSR)}$.) We denote by $t(i|s) = E_{s' \in L(s)} c^{(WSR)}(i|s')$ the expected number of times $i$ occurs in a WSR-sketch from $L(s)$. The adjusted selectivity estimator is

$$\rho_2^{(WSR)}(i) = \frac{t(i|s)}{k}.$$

If $s^{(WSR)}$ contains the keys $i_1, \ldots, i_{k'}$ ($k' \leq k$), then the probability to obtain a particular WSR sketch with $m_j + 1$ samples from $i_j$ and 0 samples from $U \setminus s^{(WSR)}$ is determined using the multinomial distribution

$$\binom{k}{m_1 + 1, \ldots, m_{k'} + 1} \prod_{h=1}^{k'} \left(\frac{w(i_h)}{w(U)}\right)^{m_h + 1}$$

$$= \frac{k! w(i_1) \cdots w(i_{k'})}{k'! w(U)^k (1 + m_1) \cdots (1 + m_{k'})} \binom{k - k'}{m_1, \ldots, m_{k'}} \prod_{h=1}^{k'} w(i_h)^{m_h}$$

The conditional expectation of the count $(m_j + 1)$ of key $i_j$ over the subspace $L(s)$ is therefore

$t(i_j|s) =$

$$= \frac{\sum_{\substack{m_1, \ldots, m_{k'} \\ \sum m_i = k - k'}} \frac{1}{\prod_{h \neq j}(1 + m_h)} \binom{k-k'}{m_1, \ldots, m_{k'}} \prod_{h=1}^{k'} w(i_h)^{m_h}}{\sum_{\substack{m_1, \ldots, m_{k'} \\ \sum m_i = k - k'}} \frac{1}{\prod_{h=1}^{k'}(1 + m_h)} \binom{k-k'}{m_1, \ldots, m_{k'}} \prod_{h=1}^{k'} w(i_h)^{m_h}}$$

$$= \frac{E_{M(k', k-k')}\left(\frac{1}{\prod_{h \in [k-k'] \setminus \{j\}}(1 + m_h)}\right)}{E_{M(k', k-k')}\left(\frac{1}{\prod_{h=1}^{k'}(1 + m_h)}\right)},$$

where the expectation is over $M(k', k - k')$, the multinomial distribution on $k'$ counts $m_1, \ldots, m_{k'}$ that sum to $k - k'$ and probabilities $p_j = w(i_j)/w(s^{(WSR)})$.

**WSRD:** The equivalence classes are according to the set $i_1, \ldots, i_k$ of the $k$ keys included in the sketch and the sum $\sum_{j=1}^{k} c^{(WSRD)}(i_j, s)$ of their multiplicities. Denote $m_j = c^{(WSRD)}(i_j, s) - 1$ for $j \in [k]$ and $\sum_{j=1}^{k} m_j = o$.

$$\frac{w(U) - w(s)}{w(U)} \binom{k + o}{m_1 + 1, \ldots, m_{k'} + 1} \prod_{h=1}^{k} \left(\frac{w(i_h)}{w(U)}\right)^{m_h + 1}$$

$$= \frac{(w(U) - w(s))(k + o)! w(i_1) \cdots w(i_k)}{k! w(U)^{o+1}(1 + m_1) \cdots (1 + m_k)} \binom{o}{m_1, \ldots, m_k} \prod_{h=1}^{k} w(i_h)^{m_h}$$

The conditional expectation of $(m_j + 1)/(k + o)$ over all WSRD sketches with set of keys $s$ and $o$ fixed is therefore

$$\frac{1}{k + o} \frac{E_{M(k, o)}\left(\frac{1}{\prod_{h \in [k] \setminus \{j\}}(1 + m_h)}\right)}{E_{M(k, o)}\left(\frac{1}{\prod_{h=1}^{k}(1 + m_h)}\right)}.$$

## 8. SCS and LCS ML estimators

Our derivations build on derivation of ML estimators for a *single* WS sketch of a weighted set [19].

We use the following Lemma, which is a consequence of the memoryless nature of the exponential distribution.

LEMMA 8.1. *[17] Consider a probability subspace of rank assignments over $J$ where the $k$ keys of smallest ranks are $i_1, \ldots, i_k$ in increasing rank order. The rank differences $r_1(J), r_2(J) - $* $r_1(J), \ldots, r_{k+1}(J) - r_k(J)$ *are independent random variables, where* $r_j(J) - r_{j-1}(J)$ $(j = 1, \ldots, k+1)$ *is exponentially distributed with parameter* $w(J) - \sum_{\ell=1}^{j-1} w(i_\ell)$. *(we formally define* $r_0(J) \equiv 0$.*)*

---

[3]The sketch can also include rank values that can be used to estimate $w(U)$ [12]. The estimate of $w(U)$ is independent of the selectivity estimators we derive. We can therefore obtain subpopulation weight estimators by multiplying the selectivity estimators with the estimate of $w(U)$. We do not need the rank values for selectivity estimation per se.

SCS ML **estimator for** $\mathbf{w}(U)$ ($U = \bigcup_{A \in \mathcal{S}} A$). Let $u_1, u_2, \ldots, u_\ell$ be the keys in $\text{SCS}_k(\mathcal{S}, r)$, sorted by increasing rank values. Let $s_i = \sum_{h=0}^{i} w(u_i)$ ($s_0 \equiv 0$).

LEMMA 8.2. *The* ML *estimator for* $w(U)$ *is the solution of the equation* $\sum_{h=0}^{\ell} \frac{1}{x - s_i} = r_{k+1}(\mathcal{S})$.

PROOF. Let $r_i \equiv r(u_i)$ ($r_0 \equiv 0$) for $i \leq \ell$ and let $r_{\ell+1} \equiv r_{k+1}(\mathcal{S})$.

Consider an equivalence class of rank assignments such that the rank order of all the keys in $U$ is as in $r$. The probability density of a rank assignment $r'$ in this class obtaining the rank values $r'_j = r_j$ for $u_j$ ($j \in [\ell]$) is

$$\prod_{i=0}^{\ell} (x - s_i) \exp(-(x - s_i)(r_{i+1} - r_i)), \quad (4)$$

where $x = w(U)$. (First note that the fixed ordering of rank values in $U$ also determines the SCS. From Lemma 8.1, the differences $r'_{i+1} - r'_i$ ($i \geq 0$) are independent exponentially distributed random variables with parameter $w(U) - s_i$.)

By taking the natural log of (4) and deriving, we obtain the estimator as the value of $x$ that maximizes this probability density. □

SCS ML **subpopulation estimator.** Consider a subpopulation $J \subseteq U$. Let $i_1, \ldots, i_m$ be the keys in $\text{SCS}_k(\mathcal{S}, r) \cap J$, in increasing rank order. Let $s_j = \sum_{h=1}^{j} w(i_h)$ ($s_0 = 0$).

LEMMA 8.3. *The solution of*

$$\sum_{i=0}^{m-1} \frac{1}{x - s_i} = r_{k+1}(\mathcal{S}).$$

*is a maximum likelihood estimator for* $w(J)$.

PROOF. Let $r_j \equiv r(i_j)$ ($r_0 = 0$).

Consider an equivalence class $R(r)$ of rank assignments such that the rank values of keys $i \in U \setminus J$ is $r'(i) = r(i)$, and the rank order induced by $r'$ on the keys of $J$ is as in $r$.

The joint probability density function in $R$ for the bottom $m$ ranks in $J$ being $r_1 < \cdots < r_m$ and the $(m+1)$st smallest rank being at least $\tau = r_{k+1}(\mathcal{S})$, as a function of $x = w(J)$ is

$$\exp(-(x - s_m)(\tau - r_m)) \prod_{h=0}^{m-1} (x - s_h) \exp(-(x - s_h)(r_{h+1} - r_h)).$$

By taking the natural logarithm and deriving, we obtain the estimator as the value of $x$ that maximizes this probability density. □

## ML Estimators that use the weights of the sets.

For data sources with explicit representation of sets, the summarization algorithm can provide the total weight of (the keys in) each set without a significant processing or communication overhead [6, 42, 17]). We derive tighter estimators that use the weight of sets.

### SCS **estimator for the weight of the intersection and union of two sets**

Let $\mathcal{S} = \{A, B\}$. Let $i_1, \ldots, i_m$ be the keys in $\text{SCS}_k(\mathcal{S}, r) \cap (A \cap B)$, in order of increasing rank values. Let $i'_1, \ldots, i'_{m'}$ be the keys in $\text{SCS}_k(\mathcal{S}, r) \cap (A \cup B \setminus A \cap B)$, in order of increasing ranks. Let $s_j = \sum_{h=1}^{j} w(i_h)$ ($s_0 = 0$) and Let $s'_j = \sum_{h=1}^{j} w(i'_h)$ ($s'_0 = 0$).

LEMMA 8.4. *The solution of*

$$\sum_{i=0}^{m-1} \frac{1}{x - s_i} - \sum_{i=1}^{m'-1} \frac{1}{w(A) + w(B) - 2x - s'_i} = 0.$$

*is an* ML *estimator for* $w(A \cap B)$.

PROOF. Consider the equivalence class $R(r)$ of rank assignments where (1) the $m$ keys of smallest $r'$ ranks from $A \cap B$ and the $m'$ keys of smallest $r'$ ranks from $A \cup B \setminus A \cap B$ are the same sets and same rank order as for $r$, and (2) that $r'_{m+1}(A \cap B) \geq r_{k+1}(\mathcal{S})$ and $r'_{m'+1}(A \cup B \setminus A \cap B) \geq r_{k+1}(\mathcal{S})$.

We compute the probability density function of the event that the $r'(i) = r(i)$ for the $m + m'$ keys in $s_k(\mathcal{S}, r)$, for $r' \in R(r)$, as a function of $x = w(A \cap B)$.

Rank values in the two disjoint sets $A \cap B$ and $A \cup B \setminus A \cap B$ are independent. The rank differences within each set are also independent.

Observe that $w(A \cup B \setminus A \cap B) = w(A) + w(B) - 2x$. We take the natural log and derive to find the value of $x$ that maximizes the probability density. □

The equation can be solved by a search on the interval $[s_{m-1}, w(A) + w(B) - s'_{m'-1}]$ as the left hand side is a monotone function of $x$.

Observe that if $x$ is the ML estimator for $w(A \cap B)$ then the ML estimator for the resemblance is $x/(w(A) + w(B) - x)$ and the ML estimator for the union is $w(A) + w(B) - x$.

### LCS **estimator for the weight of set union**

Consider a rank assignment $r$. Let $A_{i_1}, A_{i_2}, \ldots, A_{i_s}$ be the sets in $\mathcal{S}$, sorted according to $r_{k+1}(A)$. We derive ML estimator for $w(\cup_{j \in [s]} A_{i_j})$ that uses $w(A_{i_j})$ ($j \in [s]$).

For any key $x$ in the sketch of $A_{i_j}$ we know if $x \in A_{i_h}$ for all $h \geq j$. So we can apply a subpopulation ML estimator with a known total weight of [19] to the sketch of $A_{i_j}$, and get estimates for the weights of $H_j = A_{i_j} \setminus \bigcup_{h > j} A_{i_h}$ and $H'_j = A_{i_j} \cap (\bigcup_{h > j} A_{i_h})$. We have that $w(H_j) + w(H'_j) = w(A_{i_j})$. The weight of the union is $\sum_{h=1}^{s} w(H_h)$ and we obtain an estimate for the weight of the union by summing up the corresponding estimates of $w(H_h)$. (Note that $H_s = A_{i_s}$ and therefore $w(H_s)$ is known exactly.)

We can apply the same methodology to estimate the weight of a subpopulation $J \subset \bigcup_{j \in [s]} A_{i_j}$ (specified by a predicate with attribute-based conditions), by estimating $w(H_j \cap J)$, using the property that $w(H_j \cap J) + w(H'_j \cup (H_j \setminus J)) = w(A_{i_j})$.

For two subsets, $A_1$ and $A_2$ we also obtain an ML LCS estimator for $w(A_1 \cap A_2)$ using $w(A_1) + w(A_2) - \tilde{w}(A_1 \cup A_2)$, where $\tilde{w}(A_1 \cup A_2)$ is the estimate of the union. This estimate is always nonnegative, since $\tilde{w}(A_1 \cup A_2) \leq \sum_{h=1}^{s} w(A_h)$.

## 9. Empirical Evaluation

We compare our combination estimators to state of the art estimators applied to the union sketch. As a point of reference, we also include $k$-mins estimators applied to sketch of the union of $k$-mins sketches. We measure the benefit of combination estimators by their *improvement factor*, which is the ratio of average relative error of (the best) union-sketch estimator to that of the combination estimator.

**Datasets.** We used synthetic data designed to quantify and demonstrate how the quality and relative performance of the estimators depends on different parameters of the data, such as the number of relevant sets in the selection predicate and the relation between these sets. We also used the following real-life datasets that demonstrate example applications:
• Two IP packet traces of about $9 \times 10^6$ packets from gateway routers (*peering* and *campus*). These traces were partitioned into 5 consecutive time periods and we produced coordinated sketches of the set of destination IP addresses in each time period. The campus data had 3196, 2636, 2656, 2175, 2105 distinct addresses in each time period and 6830 distinct addresses overall. The peering data had 14158, 14564, 14281, 14705, 14483 distinct addresses in each time period and 37574 distinct IP addresses overall.
• The Netflix Prize [43] Data, that consists of about $1 \times 10^8$ reviews by $5 \times 10^5$ users of 17770 movies. We consider the set of reviewers of each movie as a "set," and produced coordinated sketches for these sets.

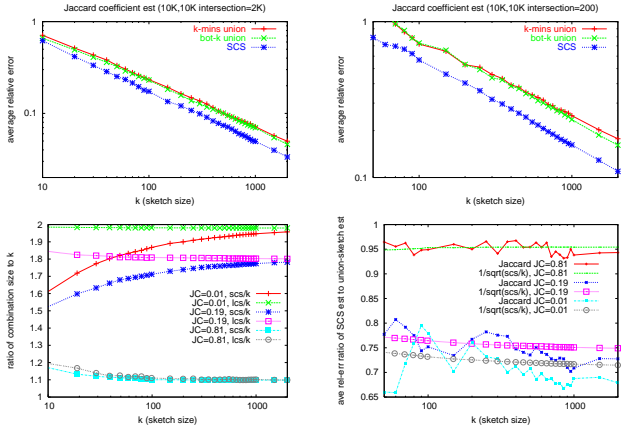

**Figure 4.** Top: Averaged relative error of different estimators for the Jaccard coefficient of two sets, each containing 10000 (uniformly weighted) keys. The size of the intersection is 2000 (left) and 200 (right). Bottom: Left: Combination sizes for 2 sets containing 10000 keys. Right: Ratio of averaged relative error of SCS to union-sketch estimators of Jaccard coefficient (with corresponding square root of the ratio of $k$ and SCS size.)

**Predicates with 2 relevant sets.** We first consider basic pairwise aggregates: The union size, intersection size, Jaccard coefficient, and Hamming distance (the difference of the sizes of union and intersection).

We use two sets $A_1$ and $A_2$ of the same sizes $|A_1| = |A_2| = 10,000$ and a varying number of common keys $|A_1 \cap A_2| \in \{200, 2000, 9000\}$ (respective Jaccard coefficients 0.81, 0.19, 0.01). We applied the RC union, $k$-mins union, and our RC SCS and RC LCS estimators for the size of the union. We applied the RC union, $k$-mins union, and our RC SCS for the size of the intersection. The intersection estimator based on inclusion exclusion and the RC LCS estimate of the union $w(A_1) + w(A_2) - \tilde{w}(A_1 \cup A_2)$ was also evaluated but it performed considerably worse than other estimators and is not shown. Hamming distance is estimated as the difference of union and intersection estimators (as the difference of unbiased estimators, this estimator is unbiased. It is also easy to show from the derivation that the estimate is always nonnegative). For the Jaccard coefficient, we applied the classic $k$-mins and bottom-$k$ union estimators of Broder [7, 6] and our SCS combination selectivity estimator (Section 7). Figure 4 shows the average relative error, over 1000 runs, of Jaccard coefficient estimators. For uniform weights and for $k$ small relative to number of keys, the relative error of the union-sketch estimators decreases proportionally to $\sqrt{k}$ and there was a proportional decrease also for the combination estimators.

The improvement factor of combination estimators is larger when the Jaccard coefficient is smaller. The intuitive reason is that smaller Jaccard coefficient means less overlap between the sets, hence less overlap between sketches, and more distinct keys in the combination that are available to the combination estimators. We relate the improvement factor to the size of the combination. Figure 4 (bottom, left) shows the ratio $\ell/k$, where $\ell$ is the average size of the combination (SCS and LCS) and $k$ is the size of the union-sketch. The figure demonstrates that the combination size is larger when the Jaccard coefficient is smaller. Figure 4 (bottom, right) shows the improvement factor and the respective $\sqrt{k/\ell}$ for our Jaccard coefficient estimators. In agreement with an analytic approximation (Section 7), we can see that the improvement factor is approximated well by $\sqrt{\ell/k}$, where $\ell$ is the combination size.

In particular, our combination estimator for Jaccard coefficient has about half the variance of union-sketch estimators [7, 6] when the two sets are almost disjoint. For the applications of identifying all similar pairs [7, 34], and on typical corpuses, with only a small fraction of pairs being similar, our estimator significantly decreases "false positives."

**Performance dependence on the number of relevant sets.** We next consider a synthetic distribution where all sets share 1000 common keys and each set has its own 5000 unique keys. This collection of sets allows us to study how the benefit of combination estimators increases with the number of sets. Figure 6 (top) shows the average relative error for estimating the size of the union of multiple (2,3,4, and 5) sets using the RC union, RC SCS, and the RC LCS estimators. The average relative error of union-size estimator applied to the sketch of the union is about $\sqrt{2/(\pi k)}$ and is about $\sqrt{2/(\pi \ell)}$ for the combination estimators. Figure 6 (bottom) shows combination size ratio to $k$. A simple calculations shows that the LCS size with $i$ sets is about $\ell = 0.2k + 0.8ik$. The SCS size ratio varies with $k$ and approaches the LCS size ratio as $k$ increases. Figure 6 also demonstrates that improvement factors are approximated well by $\sqrt{\ell/k}$, where $\ell$ is the size of the combination.

Figure 5 shows the improvement factor of SCS RC and LCS RC estimators on the destination IP addresses data sets. We estimate the total number of distinct destination IP addresses (union) and the number of common destination IP addresses (intersection) of the first $i \in \{2, 3, 4, 5\}$ time periods. The figure shows how the improvement factor of the SCS and (in particular) the LCS increases with the number of sets. The improvement factor is again approximated well by $\sqrt{\ell/k}$ (not shown).

**Performance dependence on the relation between the sets.** When sets have fewer common keys, combinations contain more keys, and combination estimators have larger improvement factors. We demonstrate this using two collections $S_1$ and $S_2$ of 5 sets each. Both collections have the same size union (49530 keys). $S_1$ contains 5 disjoint sets of 9906 keys. $S_2$ contains sets of size 29718 with 24765 keys common to all sets and 4953 exclusive keys for each set. The LCS of $\mathcal{S}_1$ contains about $5k$ keys. The LCS of $\mathcal{S}_2$ contains about $5k/3$ keys ($5/6$ of the keys in each sketch are common to all 5 sets). Figure 7 shows corresponding improvement factors of $\sqrt{5}$ for $\mathcal{S}_1$ and $\sqrt{5/3}$ for $\mathcal{S}_2$.

SCS **versus** LCS. Figures 5,6,7 show comparable performance factors (reflecting similar sizes) for the SCS and the LCS. When can we expect the SCS to be large? For $A \in \mathcal{S}$, keys in the sketch of $A \in \mathcal{S}$ are included in the SCS only if they have rank smaller than $r_{k+1}(\mathcal{S})$. Thus, when $r_{k+1}(\mathcal{S}) = \min_{B \in \mathcal{S}} r_{k+1}(B)$ is close to $r_{k+1}(A)$, most keys are included in the SCS. The SCS is large when sets have closely related distributions of $r_{k+1}(A)$ (sets have similar weights) *and* when $|\mathcal{S}|$ is smaller (see Figure 6). If there is high heterogeneity in the weight of sets in $\mathcal{S}$, the distribution of $r_{k+1}(\mathcal{S})$ is dominated by that of $r_{k+1}(A)$, where $A$ is the heaviest set in $\mathcal{S}$ – a set $B \in \mathcal{S}$ will have only about $kw(B)/w(A)$ of keys in the sketch of $B$ included in the SCS. Even with homogeneous sets the SCS is smaller when $|\mathcal{S}|$ is larger(see Figure 6).

Figure 8 shows the performance of estimators for two queries on the Netflix data set: "the number of users with at least one rating of a National Geographic title" and "the number of users with at least one rating of a movie released on or before 1930." These are estimates on the size of the union of sets. The corresponding sets of movie titles where larger (more sets than in previous datasets) and heterogeneous (high variability in number of reviewers of different titles). For the first query, there were 45 National Geographic titles with 19708 ratings by 12351 distinct reviewers. The number of ratings for each NG title varied between 93 and 1170 (mean is about 438). For the second query, there were 120 titles with release year on or before 1930. There were 117617 ratings with 53774 distinct reviewers. The number of ratings per title ranged between 54 and 12054 (mean is 980). We observe improvement factor of 3-4 of the RC LCS estimator over RC union but we also see a ratio of 1.5-2 between the relative errors of the RC SCS and the RC LCS estimates, reflecting a much smaller SCS than LCS.

Lastly, we consider the incremental effectiveness of combination samples. SCS samples (that are not included in the union-sketch) are always as effective as additional samples from the union of the sets. LCS samples (that are not included in the SCS) can be as effective, but the effectiveness decreases with heterogeneity of $\mathcal{S}$. Intuitively, consider two sets, one much larger than the other, and each contributing $k$ samples. Then samples from the smaller set (that are mostly excluded from the SCS) are much less useful to estimate properties of the union of the sets. On the other hand, if we have multiple homogeneous sets, the SCS is smaller than LCS due to the "variance" of the $k+1$st rank, but LCS samples are as effective.

$k$**-mins versus bottom-**$k$**.** "Without replacement" (bottom-$k$) estimators dominate "with replacement" ($k$-mins) estimator, but the gain is negligible with uniform weights (see Figure 4). Gain can be significant only when keys are likely to be sampled repeatedly under "with replacement" sampling [26, 17]. With uniform weights, union-sketch $k$-mins estimators performs similarly to respective union-sketch bottom-$k$ estimators *but* combination estimators typically outperform union-sketch estimators. Since combination estimators are *not applicable* to $k$-mins sketches, this suggests the use of bottom-$k$ sketches also with uniform weights.

**Weighted keys.** Improvement factor, as a function of $\ell/k$, is larger when keys are weighted. This is because variance decrease with sample size is *at least* $1/k$ (relative error decrease is *at least*

$1/\sqrt{k}$), with uniform weights exhibiting the "worst-case" decrease.

**Restricted predicates.** The demonstrated performance factor on unions and intersections of sets carries over when adding attribute based conditions to the predicate. This is because also with added conditions, the combination contains proportionally more keys than the union sketch. Examples of attribute-based conditions (on IP addresses) is to restrict the query to blacklisted addresses or addresses that belongs to a particular Autonomous System or (on Netflix-like data) to reviewers from a certain gender or zip-code.

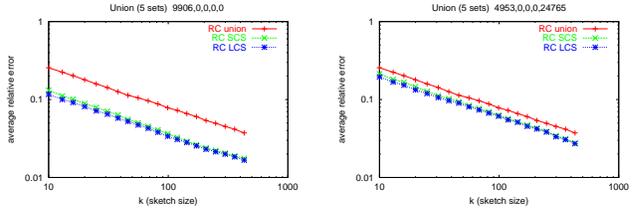

**Figure 7.** Relative error of RC union, RC SCS, and RC LCS estimators on the size of the union of 5 sets. Size of the union is 49530. Left: 5 disjoint sets of size 9906. Right: 5 sets with 24765 common keys to all 5 and 4953 exclusive keys in each set.

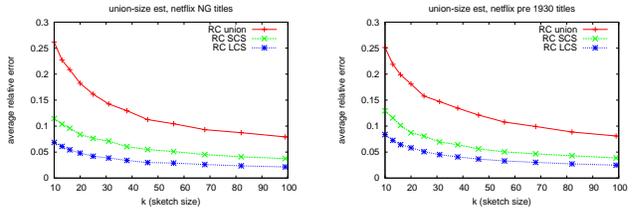

**Figure 8.** Relative error of RC union, RC SCS, and RC LCS estimators on "number of distinct reviewers of National Geographic titles" and "number of distinct reviewers of titles released on or before 1930." (Netflix data set)

## 10. Related work

**Sample-based coordinated sketches.** Coordinated samples of multiple sets, based on keys "retaining" the same "random draw" across sets, are extensively used as a way to maximize or minimize sample overlap [5, 44, 45, 47, 48] or to facilitate (approximate) aggregations over distinct keys [12, 30, 31, 7, 16, 17, 19]. Sample-based coordinated sketches where used with size-$k$ samples with replacement ($k$-mins sketches) [12, 7, 16], size-$k$ samples without replacement (bottom-$k$/order samples) [44, 45, 47, 12, 17, 19], and Poisson sampling [5, 30, 31].

Bottom-$k$ sampling [46, 33, 44, 45, 47, 26, 17, 19] has the advantage (over Poisson and with-replacement sampling) of fixed sample size and tighter estimators.

**Multiple-set aggregates.** The union-sketch reduction was used with both $k$-mins and bottom-$k$ sketches [12, 7, 6, 2, 25] and we are not aware of a better estimator over $k$-mins sketches. The only previous work we are aware of that leveraged combinations of bottom-$k$ sketches is [38], but they only derive ML estimators that are biased and applicable only to WS bottom-$k$ sketches. Multiple-set aggregates over Poisson samples are approximated by producing a Poisson sample of the union of the sets [30, 31]. Poisson samples have the disadvantage of variable sample size. With our combination estimators, coordinated bottom-$k$ sketches dominate other sampling-based sketching methods by providing both fixed sample size and tighter estimators.

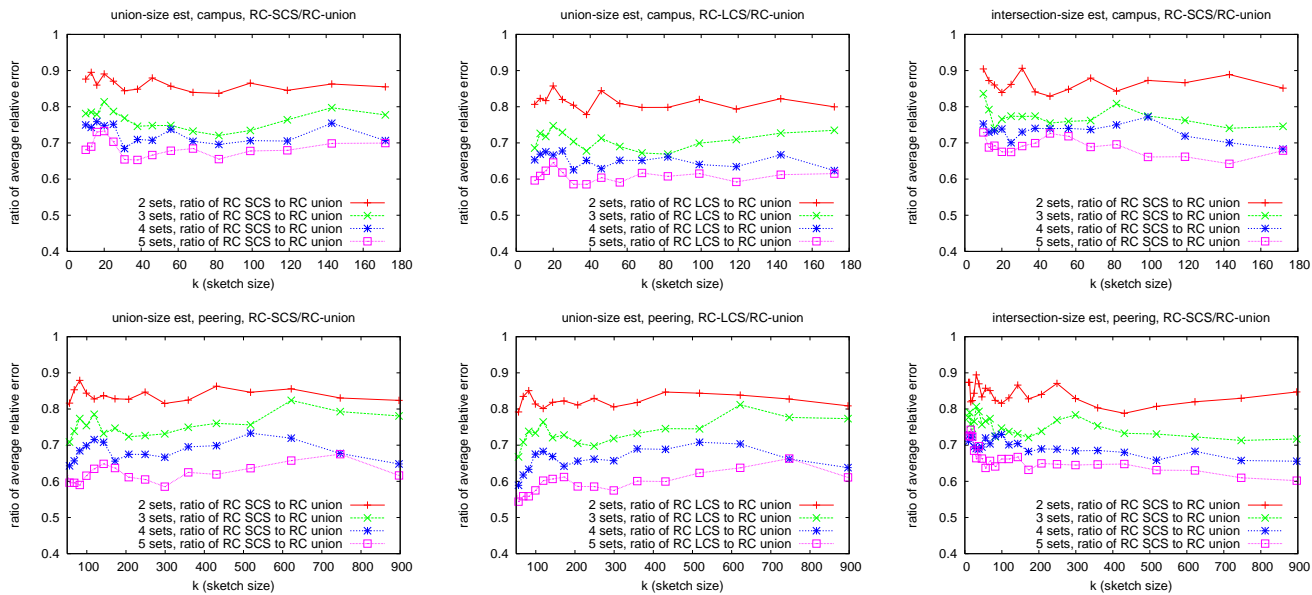

**Figure 5.** Ratio of the average relative error to that of the RC union estimator (inverse improvement factor) as a function of $k$. Applied to sketches of destination IP addresses in $i \in \{2, 3, 4, 5\}$ consecutive time periods (Top: *campus* data set, Bottom: *peering* data set). Left and Middle: RC SCS and RC LCS estimate of the union of the first $i$ time periods. Right: RC SCS estimate of the intersection of the first $i$ periods.

**Coordinated sketches that are not sample based.** A strength of sampling-based coordinated sketches is the generality of the selection predicates combined with a tunable and potentially very small summary size. Methods that are not sample-based include bloom filters [4] and variants [28] that have the drawback that summary size grows linearly with the size of the corpus. Other methods, such as Charikar's simhash [10], produce tunable small-size summaries [39, 29, 10, 11, 50, 23, 34, 37, 40]. These methods are very effective for some tailored goals, such as pairwise similarity measures between sets [34], but have inherent limitations: Since the summary does not retain keys' identifiers or meta-data, there is no support for predicates with attribute-based conditions. For example, in a market basket data set, where baskets are "keys" and goods are "sets," we can estimate the association "purchase of beer implies purchase of diapers", using the ratio of the number of baskets with beer and diapers (size of the intersection) and the number of baskets with beer. The more refined query where the selection is restricted to consumer/basket segments (such as "female consumer," "basket contains at most 12 goods," or "paid in cash."), however, can not be supported. Furthermore, only a limited set of membership-based conditions is supported and inherently these methods do not provide a "representative sample" of keys that satisfy the predicate.

A recent sampling/summarization scheme, *varopt*, minimizes the sum of variances of sets of any fixed size [13]. We do not know how to apply it to produce coordinated sketches.

This paper expands on a 6-page exposition [18] and on a conference version [20].

## 11. Conclusion

Sketches based on coordinated samples are a classic summarization method for datasets modeled as a collection of sets over a ground set of keys. The sketch of each set is a weighted sample of the keys with some auxiliary information. This powerful model covers a wide range of applications and sample-based sketches facilitate a much wider class of approximate queries than other sketching methods.

We propose novel unbiased estimators for multiple-set weight and selectivity aggregates over coordinated bottom-$k$ sketches. Our *combination* estimators outperform the existing union-sketch estimators by using more samples present in the sketches of the sets relevant to the query. We quantify the advantage of combination estimators over union-sketch estimators through an extensive empirical evaluation. Our evaluation suggests that combinations estimators applied when the combination has average size of $\ell$ (has $\ell$ distinct keys) perform *comparably* to estimators applied to a size-$\ell$ union-sketch (derived from coordinated bottom-$\ell$ sketches). In particular, we can expect $\ell/k$ factor reduction in variance ($\sqrt{\ell/k}$ reduction in estimation error) for uniform weights (distinct values count) and a larger factor for skewed distributions. The size $\ell$ of a combination is between $[k, t*k]$, where $t$ is the number of relevant sets. Combination size is larger when there are more sets, when sets have fewer common keys, and when sets have homogeneous weights. Our evaluation, which includes natural queries on real data sets demonstrate typical 25%-4 fold reduction in estimation error.

With our combination estimators, coordinated bottom-$k$ sketches, that have the advantage (over other sample-based coordinated sketches)

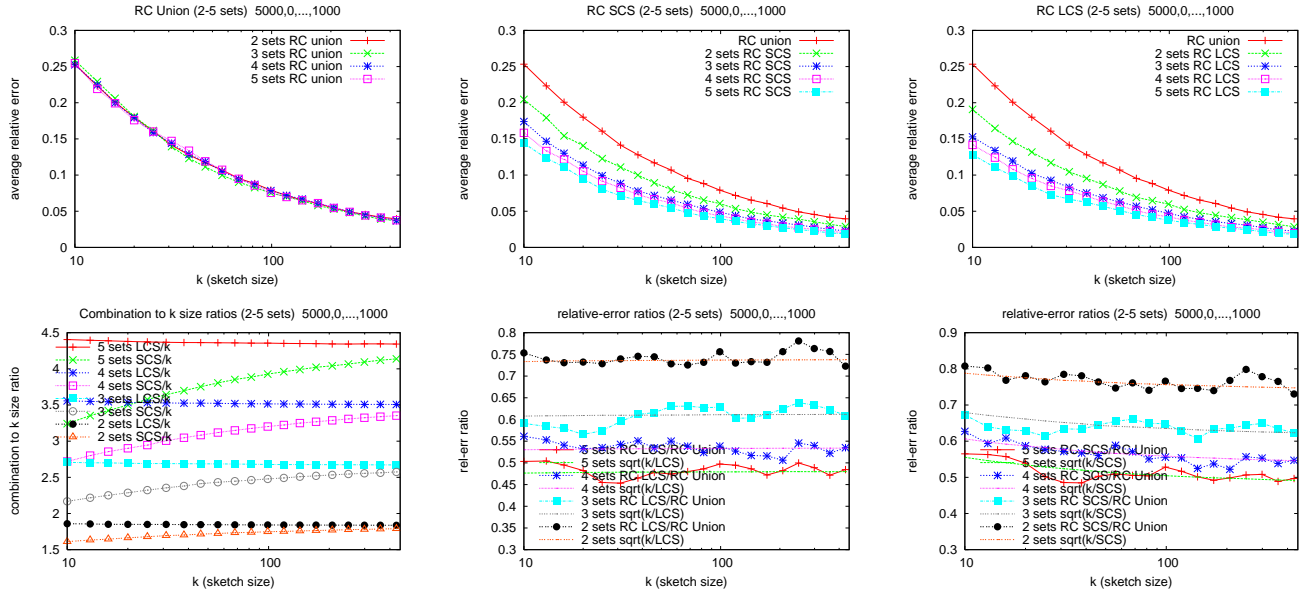

**Figure 6.** Top: Relative error of RC union, RC SCS, and RC LCS estimators on the size of the union of 2,3,4, and 5 sets of size 5000 each with intersection of size 1000. Bottom: size ratios of combinations to $k$ (left) and relative error ratios for LCS (middle) and SCS (right).

of fixed-size and scalable algorithms, dominate other coordinated sampling methods by also providing tighter estimators. We therefore expect them to become the method of choice.